\documentclass[pra,twocolumn,floatfix,superscriptaddress,amsmath,aps,longbibliography]{revtex4-1}
\pdfoutput=1
\usepackage[colorlinks,plainpages=false,linkcolor=blue,urlcolor=blue,citecolor=blue,pdfpagemode=UseNone,pdfstartview=FitBH]{hyperref}
\usepackage{dcolumn,graphicx,color,booktabs,microtype,afterpage}
\makeatletter\renewcommand{\fnum@figure}[1]{\figurename~\thefigure.}\makeatother
\makeatletter\renewcommand{\fnum@table}[1]{\tablename~\thetable.}\makeatother
\newcommand{\YYbAl}{Yb:YAlO$_3$}
\newcommand{\YbAl}{YbAlO$_3$}
\newcommand{\bS}{\mbox{\boldmath$S$}}
\newcommand{\bM}{\mbox{\boldmath$M$}}

\begin{document}

\title{Experimental observation of magnetic dimers in diluted Yb:YAlO$_3$}
\author{S.~E.~Nikitin}
\affiliation{Max Planck Institute for Chemical Physics of Solids, D-01187 Dresden, Germany}
\affiliation{Institut f{\"u}r Festk{\"o}rper- und Materialphysik, Technische Universit{\"a}t Dresden, D-01069 Dresden, Germany}
\author{Tao~Xie}
\affiliation{Neutron Scattering Division, Oak Ridge National Laboratory, Oak Ridge, Tennessee 37831, USA}
\author{A.~Podlesnyak}
\email[{Correspondence to: }]{podlesnyakaa@ornl.gov}
\affiliation{Neutron Scattering Division, Oak Ridge National Laboratory, Oak Ridge, Tennessee 37831, USA}
\author{I. A. Zaliznyak}
\email{zaliznyak@bnl.gov}
\affiliation {Condensed Matter Physics and Materials Science Division, Brookhaven National Laboratory, Upton, NY 11973, USA}

\begin{abstract}
We present a comprehensive experimental investigation of Yb magnetic dimers in Yb$_{0.04}$Y$_{0.96}$AlO$_3$, an Yb-doped Yttrium Aluminum Perovskite (YAP) YAlO$_3$ by means of specific heat, magnetization and high-resolution inelastic neutron scattering (INS) measurements. In our sample, the Yb ions are randomly distributed over the lattice and $\sim 7$\% of Yb ions form quantum dimers due to nearest-neighbor antiferromagnetic coupling along the $c$-axis. At zero field, the dimer formation manifests itself in an appearance of an inelastic peak at $\Delta \approx 0.2$~meV in the INS spectrum and a Schottky-like anomaly in the specific heat. The structure factor of the INS peak exhibits a cosine modulation along the $L$ direction, in agreement with the $c$-axis nearest-neighbor intra-dimer coupling. A careful fitting of the low-temperature specific heat shows that the excited state is a degenerate triplet, which indicates a surprisingly small anisotropy of the effective Yb-Yb exchange interaction despite the low crystal symmetry and anisotropic magnetic dipole contribution, in agreement with previous reports for the Yb parent compound, YbAlO$_3$~\cite{Wu2019,WuPRB2019}, and in contrast to Yb$_2$Pt$_2$Pb \cite{Wu2016,Gannon2019}. The obtained results are precisely reproduced by analytical calculations for the Yb dimers.
\end{abstract}

\maketitle{}

\section{Introduction}
Rare Earth based systems have recently been widely considered among the leading candidates for quantum information storage and processing \cite{Bertaina2007,Simon2010,Thiel2011,Awschalom2018,Kunkel2018}. In rare earths, strong relativistic spin-orbit coupling ties the spin and the orbital angular momentum together, so they form a multiplet of total angular momentum, ${\cal J}$, often with large ${\cal J}$. In the presence of crystal electric field (CEF) the ($2{\cal J}+1$)-fold degeneracy of the orbital multiplet is lifted and the states of the rare-earth ion are split in energy. In the case of Kramers ions, such as Ce$^{3+}$, Nd$^{3+}$, Dy$^{3+}$, Er$^{3+}$, Yb$^{3+}$, the CEF level splitting retains Kramers twofold degeneracy and the lowest doublet can be well separated in energy from the rest of the multiplet, forming quantum degree of freedom that can be described by an effective spin-1/2. These quantum doublets are protected by time-reversal symmetry and can form entangled spin states. Their peculiar coupling with external electromagnetic filed is governed by the angular momentum, ${\cal J}$, conservation, thus offering a route towards protected quantum states in rare earth ion magnets, which are suitable for solid state quantum information processing and storage.

In the case of 4$f$ ions with non-zero orbital angular momentum, the CEF splitting is typically only a modest perturbation to the dominant spin-orbit splitting. Consequently, in an ion such as Yb$^{3+}$, there are a number of ${\cal J} \rightarrow {\cal J} \pm 1$ optical transitions at slightly different wavelengths, which can be individually addressed by photons.
The compatibility with optical technology makes Yb and sister rare earths (Ce, Nd, Er, Dy) attractive for incorporating in optics-based quantum computation and information storage schemes with coherent and entangled photons \cite{Thiel2011,Awschalom2018,Kunkel2018,Olmschenk2009,Lim_PRB2018}. Rare-earth based insulating crystals such as Yb:YAP (Yb-doped Yttrium Aluminum perovskite, Yb$_{x}$Y$_{1-x}$AlO$_{3}$), Yb:YAG (Yb-doped Yttrium Aluminum garnet, Yb$_{x}$Y$_{3-x}$Al$_{5}$O$_{12}$), or Nd:Y$_{2}$SiO$_{5}$ are well known laser materials with high quantum efficiencies, long relaxation times and narrow optical transition lines in the near infrared range overlapping with telecom wavelengths \cite{Fagundes2007,Boulon2008}. Quantum repeaters \cite{Simon2010}, teleportation between distant condensed matter qubits \cite{Olmschenk2009}, optical preparation of coherent dark states \cite{XiaPRL2015}, quantum photon teleportation and storage in a quantum memory \cite{Bussieres2014}, and quantum storage of entangled photons \cite{Saglamyurek2015} and an all-optical retrieval \cite{Zhong2017}, were all recently demonstrated in rare-earth laser materials.

Recently, magnetic quantum doublet degree of freedom in materials with $4f$ ions with odd number of $f$-electrons, such as Ce$^{3+}$, Nd$^{3+}$, Yb$^{3+}$, etc., which is described by an effective spin-1/2, came to prominence in the context of novel quantum magnets. Among these are two-dimensional triangular-lattice systems \cite{paddison2017continuous, baenitz2018naybs}, three-dimensional pyrochlore-lattice spin-liquids, e.g. Ce$_2$Zr$_2$O$_7$~\cite{gao2019experimental}, quantum dimer magnet Yb$_2$Si$_2$O$_7$~\cite{hester2019novel}, one-dimensional spin-chain antiferromagnets (AFM) Yb$_2$Pt$_2$Pb~\cite{Wu2016,Gannon2019}, YbFeO$_3$~\cite{Nikitin2018} and YbAlO$_3$~\cite{Wu2019, WuPRB2019}.
Traditionally, the physical realization of such systems were found among the $3d$-electron (mostly Cu$^{2+}$) materials, such as frustrated spin-liquid candidate herbertsmithites (see Ref.~\cite{norman2016colloquium} and references therein), two-dimensional Heisenberg AFM PHCC \cite{Stone2006} and Cu(DCOO)$_2\cdot$4D$_2$O~\cite{dalla2015fractional}, one-dimensional spin-chain materials SrCuO$_2$ and Sr$_2$CuO$_3$~\cite{kim1996observation, motoyama1996magnetic,Zaliznyak_PRL2004,Walters_NaturePhysics2009}, etc. These materials, however, usually have significant disadvantages, which complicate their investigations: (i) a small magnetic moment; (ii) a very strong exchange interaction, typically $J > 10$~meV, which means that one has to apply an exceptionally high magnetic field to influence the ground state. In rare earth based systems, on the other hand, the overlap of the $4f$ electronic wavefunctions is small and effective spin interactions are often quite weak, making detailed magnetic field studies possible \cite{Wu2019,WuPRB2019,Wu2016,Gannon2019}.

YbAlO$_3$, an end member of Yb:YAP family of materials was shown to be a good realization of $S = 1/2$ one-dimensional antiferromagnet, which exhibits spinon confinement-deconfinement transition and quantum criticality in different regions of its phase diagram~\cite{Wu2019}. Its physics can be reasonably well described by a combination of nearest-neighbor intrachain exchange interaction along the $c$-axis and weak in-plane interchain dipolar interaction, which stabilizes the magnetic order below $T_{\mathrm{N}} = 0.88$~K.
It is worth noting that this is an unexpected result because the crystal structure of YbAlO$_3$ does not naturally imply any ``special'' chain direction (this is similar to what was found in Yb$_2$Pt$_2$Pb~\cite{Wu2016}). Moreover, while the symmetry of Yb site is very low ($C_s$) and the CEF environment induces an exceptionally high magnetic anisotropy of the ground state doublet ($g_z >> g_{xy}$), the analysis of INS spectra and critical exponents has shown that the intrachain exchange interaction is very isotropic, $0.88J_{xy} \lesssim J_{z} \lesssim J_{xy}$.

We note that low-dimensional $S = 1/2$ magnets remain at the cutting edge of the solid state physics for the last decades. Depending on dimension of the lattice they can host exotic ground states and excitations. For instance, two-dimensional triangular AFM is predicted to host entangled quantum spin-liquid ground state. One-dimensional spin models show no magnetic ordering down to zero temperature and exhibit fractionalized spinon excitations. Even the simplest toy model -- a dimer consisting of two spins coupled by an AFM interaction has an entangled ground state, $\psi = \frac{1}{\sqrt{2}}\lvert \uparrow\downarrow - \downarrow\uparrow\rangle$, which has no clear analogy in classical physics and presents a useful two-qubit model for quantum information science (QIS). Hence, studies of these systems presents significant interest.

In this work, we investigate the magnetic Hamiltonian of Yb-Yb effective spin interactions in \YYbAl by measuring the diluted sample, Yb$_{x}$Y$_{1-x}$AlO$_{3}$ ($x \approx 0.04$). In this material, the Yb ions are statistically distributed over the Y sites. If two Yb ions would occupy nearest neighbor positions along the $c$-axis, they can form a magnetic dimer. By studying the dimer physics one can refine the parameters of Yb-Yb exchange interactions with very high precision. Thereby, here we aim to answer two main questions: (i) Are there magnetic dimers in \YYbAl? (ii) What is the exchange Hamiltonian for the dimers?

\section{Experimental details}

The YAlO$_3$ single crystal doped with (nominal) 5\% of Yb$^{3+}$ (Yb:YAlO$_3$) was procured commercially from Scientific Materials \cite{SM}.

Magnetic measurements were carried out using MPMS3-VSM with magnetic fields up to 7~T and temperatures down to 1.8~K. Specific heat measurements of $m = 11.5$~mg sample of \YYbAl\ were carried out using a commercial PPMS from Quantum Design with $^{3}$He option. Magnetic field was applied along the easy $a$-axis.

Neutron scattering measurements were performed at the time-of-flight Cold Neutron Chopper Spectrometer (CNCS)~\cite{CNCS1,CNCS2} at the Spallation Neutron Source at Oak Ridge National Laboratory. Data were collected with a single crystal \YYbAl\ sample with mass around 0.6~g, which was aligned in the $(0KL)$ scattering plane. Triton - Cryofree dilution refrigeration system from Oxford Instruments was used to cool the sample down to $T=0.1$~K. The data were collected using fixed incident neutron energies of $E_{\mathrm{i}}=1.55$~meV ($\lambda_{\mathrm{i}}=7.27$~{\AA}) and $E_{\mathrm{i}}=3.32$~meV ($\lambda_{\mathrm{i}}=4.96$~{\AA}) resulting in a full-width-at-half-maximum energy resolution at the elastic position of 0.04~meV and 0.11~meV, respectively. The measurements were performed using the rotating single crystal method.
All time-of-flight data-sets were combined to produce a four-dimensional scattering-intensity function $I(\mathbf{Q},\hbar\omega)$, where $\mathbf{Q}$ is the momentum transfer and $\hbar\omega$ is the energy transfer.
The software packages~\textsc{Horace}~\cite{Horace} and~\textsc{MantidPlot}~\cite{Mantid} were used for data reduction and analysis.

\section{Results and analysis}

\subsection{Origin of the dimers}

\begin{figure}[tb!]	
\center{\includegraphics[width=1.\linewidth]{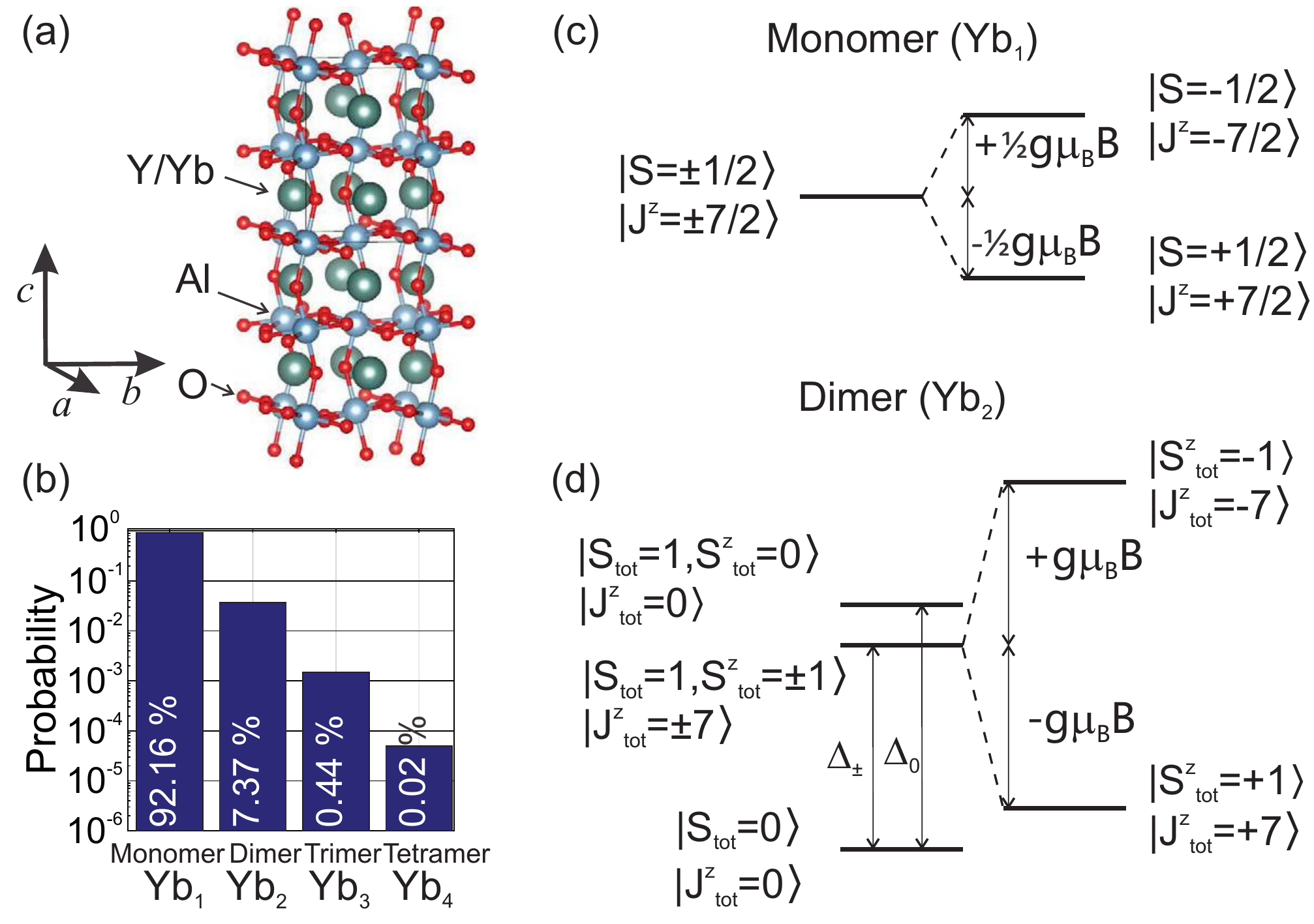}\vspace{-0pt}}
\caption{~Crystal structure and energy levels of Yb dimers in \YYbAl. (a) Orthorhombic perovskite $Pbnm$ crystal structure of Y$_{1-x}$Yb$_x$AlO$_3$ where Yb form chain segments along the $c$-axis. (b) Calculated probability for Yb ions to form $n$-mers of different length; the ratio $\frac{n\cdot P(\text{Yb}_{n+1})}{(n+1)\cdot P(\text{Yb}_{n})}$ is constant and equal to the concentration of Yb, $x \approx 0.043$. (c) Effective spin-1/2 energy levels of the ground state doublet of an isolated Yb$^{3+}$ ion and (d) splitting of the Yb dimer levels in the presence of XXZ interaction, Eq.~\eqref{XXZ_1}, and magnetic field, $B$.}
\label{Fig1:dimer_levels}\vspace{-12pt}	
\end{figure}

The orthorhombically distorted perovskite crystal structure of Yb:YAlO$_3$ ($Pbnm$, $a = 5.180$~\AA, $b = 5.330$~\AA, $c = 7.375$~\AA\ at room temperature) is illustrated in Fig.~\ref{Fig1:dimer_levels}(a) \cite{Wu2019,WuPRB2019,Buryy2010,Noginov2001}.
In pure \YbAl, $4f$ orbitals of Yb ions have non-negligible overlaps only along the $c$-axis~\cite{Wu2019}, meaning that Yb magnetic moments form chains where each Yb is only coupled with its two nearest neighbors. This is  due to the hexagonal, 6-fold symmetry of $L=3$ orbital wave function of a single unpaired electron in $4f^{13}$ shell of Yb$^{3+}$ and is similar to the case of Yb$_2$Pt$_2$Pb~\cite{Wu2016}.

As we show below, in our Yb:YAlO$_3$ sample $\approx 96$~\% Yb ions are substituted by nonmagnetic Y ions (the nominal concentration is 95~\%). Assuming that the Yb ions are randomly distributed over the sample, we can calculate the probability to find Yb multimer of length $n$ both analytically and numerically, using Monte-Carlo simulation.

If we consider a single Yb ion, the probability that it will have only Y neighbors along the chain direction is equal to $P_1 = (1-x)^2$ (throughout the text $x$ represents relative concentration of Yb ions in the sample, Yb$_x$Y$_{1-x}$AlO$_3$). It is easy to show that for dimer $P_2 = x(1-x)^2$, for trimer $P_3 = x^2(1-x)^2$ and finally for $n$-mer $P_n = x^{n-1}(1-x)^2$. The total number of all $n$-mers in the chain, $\sum_i^{\infty } P_i = 1-x$, counts the number of Y ions and the probability of Yb ion to be in an $n$-mer is $nx^{n-1}(1-x)^2$. Thereby, in \YYbAl\ 92~\% of all Yb ions occupy single sites, 7.4~\% form magnetic dimers and $\sim0.5$~\% form longer structures. To crosscheck these conclusions classical Monte-Carlo calculations were used. We generated a chain with length of $10^8$ ions with $4\cdot10^6$ randomly distributed impurity atoms and calculated number of single Yb, dimers, etc. Figure~\ref{Fig1:dimer_levels}~(b) shows the resulting histogram, which is in perfect agreement with the analytical results.

\subsection{Effective spin Hamiltonian and level splitting}\label{Sec:Hamiltonian}
The crystal electric field (CEF) splits the ${\cal J}=7/2$ Yb spin-orbit octet in \YbAl\ into four Kramers doublets. The lowest doublet corresponds to nearly pure ${\cal J}^{z} = \pm 7/2$ and is separated by a substantial energy gap, $\Delta E/k_{\mathrm{B}} \sim 300$~K ($k_{\mathrm{B}}$ is the Boltzmann constant), from the first excited doublet \cite{Wu2019,Wu2016}. This leads to an effective spin-1/2 description of the low-energy and low-temperature behavior for $T \ll \Delta E/k_{\mathrm{B}}$ where the higher doublets are not populated. The effective Hamiltonian is that of an $S=1/2$ XXZ chain~\cite{Wu2016}, which for a dimer reduces to,
\begin{equation}
H = J \left( S_{1}^{x}S_{2}^{x} + S_{1}^{y}S_{2}^{y} + \Delta S_{1}^{z}S_{2}^{z} \right).
\label{XXZ_1}
\end{equation}
Here, $J$ is the effective exchange coupling and $\Delta$ is its anisotropy. In  Yb$_2$Pt$_2$Pb, it was found that the effective Hamiltonian has Ising anisotropy, $\Delta \sim  2 - 3$ and $J \sim 0.1 - 0.2$~meV~\cite{Wu2016}. In the case of \YbAl, Ising gap is absent in the magnetically disordered state above $T_{\mathrm{N}}$ and isotropic XXX Hamiltonian with $J = 0.2$~meV adequately describes magnetic excitations, albeit an XY-type anisotropy, $\Delta \lesssim 1$, could not be definitively ruled out~\cite{Wu2019,WuPRB2019}.

Measuring the excitation spectra of a diluted system dominated by dimers provides a simple and direct way to probe the Hamiltonian, such as Eq.~\eqref{XXZ_1} \cite{furrer2013magnetic}. The dimer Hamiltonian~\eqref{XXZ_1} is straightforwardly diagonalized in terms of the total effective dimer spin, $\bS_{\mathrm{tot}} = \bS_1 + \bS_2$, and its $z$-component, $S_{\mathrm{tot}}^z$,
\begin{equation}
H = J \left[ \frac{1}{2} \left( \bS_{\mathrm{tot}} \right)^{2} + \frac{\Delta - 1}{2} \left( S_{\mathrm{tot}}^z \right)^{2} - \frac{2\Delta + 1}{4} \right].
\label{XXZ_2}
\end{equation}
In the presence of anisotropy, $\Delta \neq 1$, the $S_{\mathrm{tot}} = 1$ triplet splits into a $S_{\mathrm{tot}}^z = \pm 1$ doublet and a $S_{\mathrm{tot}}^z = 0$ singlet. For $\Delta < 1$, the doublet gap, $\Delta_{\pm} = J \frac{1 + \Delta}{2}$, is smaller than the gap to a singlet, $\Delta_{0} = J$ (Fig.~\ref{Fig1:dimer_levels}).

In magnetic field, $B$, directed along the quantization $z$-axis, the $S_{\mathrm{tot}}^z = \pm 1$ doublet splits according to [Fig.~\ref{Fig1:dimer_levels}~(d)],
\begin{eqnarray}
\Delta_{\pm} (B) & = &\Delta_{\pm} \mp g \mu_B B = J \frac{1 + \Delta}{2} \mp g \mu_B B \nonumber \\
\Delta_{0} (B) & = &\Delta_{0} = J .
\label{Zeeman_splitting_lowB}
\end{eqnarray}
Here, Lande $g$-factor relates magnetic moment along the field direction with the corresponding effective spin-1/2 component, $M^z = g \mu_B S^z$. Above the critical field, $B_{\mathrm{c}} = \Delta_{\pm}/(g\mu_{\mathrm{B}})$, the gap to $\Delta_{+}$ level closes and it becomes the ground state. For $B>B_{\mathrm{c}}$,
\begin{eqnarray}
\Delta_{-} (B) & = &\Delta_{\pm} + 2g \mu_{\mathrm{B}} (B - B_{\mathrm{c}}) \nonumber \\
\Delta_{0} (B) & = &\Delta_{0} + g \mu_B (B - B_{\mathrm{c}}) \\
\Delta_{0,0} (B) & = & g \mu_{\mathrm{B}} (B - B_{\mathrm{c}}) \nonumber ,
\label{Zeeman_splitting_highB}
\end{eqnarray}
where $\Delta_{0,0}$ is the gap between the new $\Delta_{+}$ ground state and the $S_{\mathrm{tot}} = 0$ singlet.

While all three excited states contribute to entropy and the heat capacity both below and above $B_{\mathrm{c}}$, the dipole selection rules, $\Delta {\cal J}^z = 0, \pm1$ (angular momentum conservation), only allow magnetic transitions between the singlet, $|S_{\mathrm{tot}} = 0 \rangle$, and the $|S_{\mathrm{tot}} = 1, S^z_{\mathrm{tot}} = 0 \rangle$ component of the triplet. Hence, only this, $\Delta {\cal J}^z = 0$ transition contributes to neutron scattering cross-section. The energy of this transition is $\Delta_0 = J$, independent of the anisotropy and magnetic field, but its intensity fades above $B_{\mathrm{c}}$ where the transition becomes thermally activated. At zero magnetic field, the intensity of this singlet-triplet transition is determined by the level populations \cite{Leuenberger_PRB1984,Zheludev_PRB1996} and also decreases with the increasing temperature, according to the temperature prefactor,
\begin{equation}
p(T) = \frac{1}{1 + 3 e^{-\frac{\Delta_0}{k_B T}}} ,
\label{Eq:T_prefactor}
\end{equation}
which changes from 1 at $T = 0$ to $\simeq 0.25$ in the equipartition regime at high temperature.

\subsection{Specific heat}\label{Sec:SpecHeat}

\begin{figure}[tb!]	
\center{\includegraphics[width=1\linewidth]{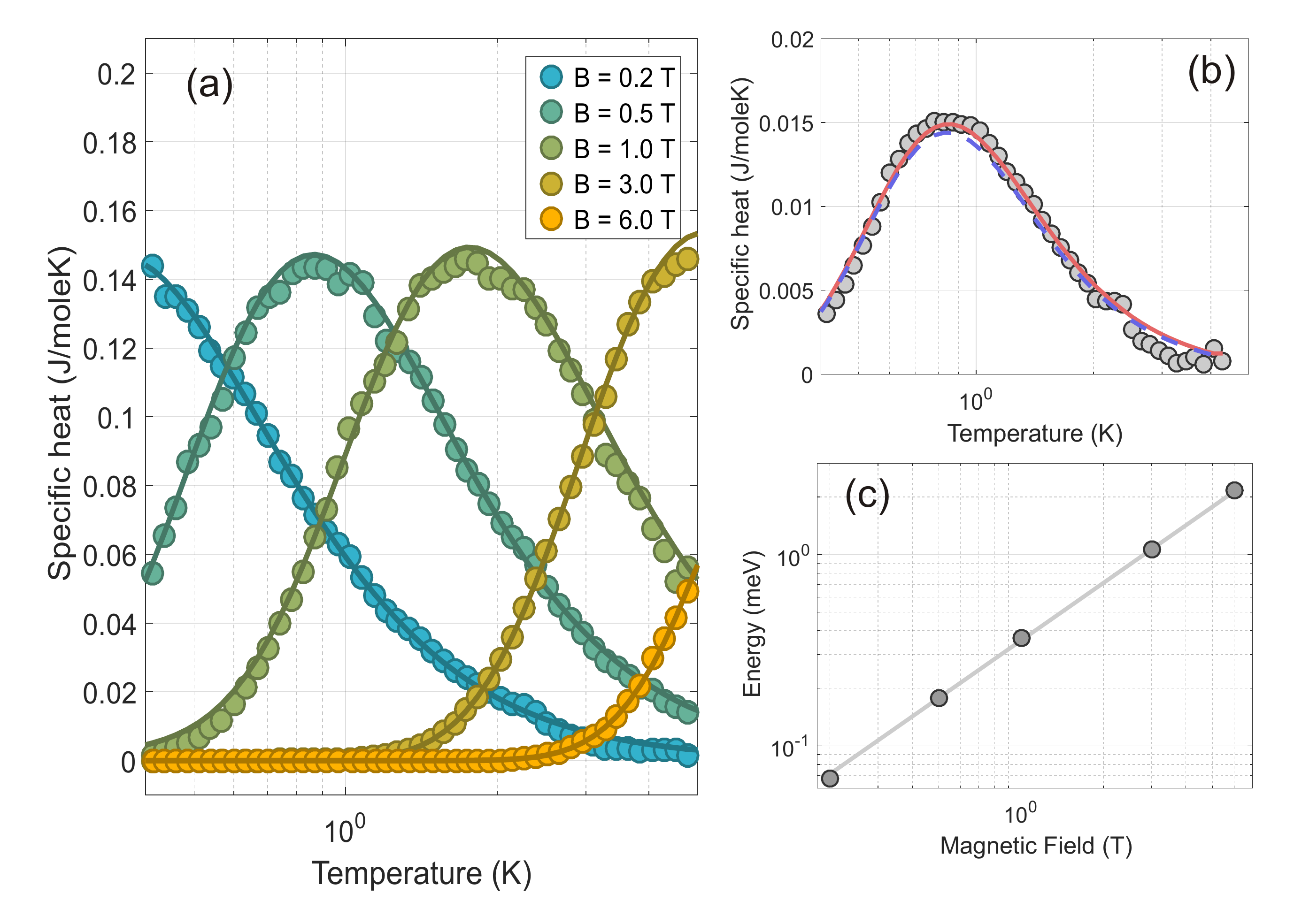}\vspace{-10pt}}
\caption{~Low-temperature specific heat of Yb:YAlO$_3$.
 		 (a)~Temperature dependences of specific heat measured at different magnetic field ($B \parallel [100]$) along with the results of fitting using Eq.~(\ref{Eq:Specific_heat_3}).
 		 (b)~Specific heat taken at 0~T. Note that the $y$-scale is an order of magnitude smaller compare to panel (a). Blue and red lines show calculated specific heat of Yb subsystem taking into account only the dimer and the dimer plus trimer contributions using Eq.~\eqref{Eq:Specific_heat_1} and Eq.~\eqref{Eq:Specific_heat_2}, respectively.
 		 (c)~Energy splitting for Yb$_1$ subsystem as a function of magnetic field extracted from the specific heat measurements and its linear fitting $E(B) = g\mu_{\mathrm{B}}BS$.}
\label{SpecificHeat}\vspace{-12pt}
\end{figure}
We start exposition of our experimental results with the low-temperature magnetic specific heat measured at different magnetic fields, which is presented in Fig.~\ref{SpecificHeat}. The phonon contribution was fitted with the standard Debye power-law dependence, $C \propto T^3$, and subtracted from the data. The curve taken at 0~T is shown in Fig.~\ref{SpecificHeat} (b) and one can see a clear Schottky-like anomaly with a maximum at $T \approx 0.85$~K. Figure~\ref{SpecificHeat} (a) presents the data measured at different magnetic fields up to 6~T. One can see that: (i) a new, strong, Schottky-like peak appears, which arises from the contribution of individual Yb moments, Yb$_1$, to the specific heat and significantly exceeds the zero-field signal in intensity; (ii) magnetic field gradually shifts the peak due to the Zeeman effect.

To describe the measured specific heat of \YYbAl\ we use a standard equation for specific heat of a discrete $p$-level system,
\begin{eqnarray}
C_{\mathrm{n-mer}}(T) =  R \frac{d}{d T} \Big( \frac{1}{\mathcal{Z}} \sum_{i = 1}^{p = 2^n} \ E_{{i}}e^{-\frac{E_{{i}}}{k_{\mathrm{B}} T} }\Big),
\label{MultiSchottky}
\end{eqnarray}
where $R$ is universal gas constant, $E_i$ are the energy levels, $\mathcal{Z}$ is the partition function and summation runs over all states of the system ($p = 2^n$ for an $n$-mer subsystem).

At low magnetic field, $B < B_{\mathrm{c}}$, the splitting of the Yb doublet is described by Eqs.~(\ref{Zeeman_splitting_lowB}). Note that at zero field Yb monomers have a degenerate doublet ground state and do not contribute to the specific heat. Thus, the signal can be considered as a specific heat of dimers plus a weak contribution of longer $n$-mers with $n>2$. At $B = 0$ the dimer level splitting scheme is a doublet at $\Delta_{\pm}$ and a singlet at $\Delta_{0}$ [see Fig.~\ref{Fig1:dimer_levels}(d)]. In order to evaluate these parameters, as well as the actual concentration of Yb in our sample we used Eq.~(\ref{MultiSchottky}) to fit the $B=0$ data, taking into account the dimer concentration [Fig.~\ref{SpecificHeat} (b)],
\begin{eqnarray}
C_{\mathrm{Mag}}(T) = x^2 (1-x)^2 C_{\mathrm{Dimer}}(T).
\label{Eq:Specific_heat_1}
\end{eqnarray}
The analysis revealed $\Delta_{\pm} = 0.2146(9)$~meV, $\Delta_{0} = 0.2142(9)$~meV, and concentration $x = 0.0439(3)$. Surprisingly, within rather high precision of our fit we did not observe any appreciable splitting of the triplet. Hence, we find $\Delta_{\pm} \approx \Delta_{0}$, implying effective isotropic Heisenberg exchange interaction, $\Delta \approx 1$ in Eq.~\eqref{XXZ_1}, in agreement with what was inferred in Ref.~\onlinecite{Wu2019} from the analysis of magnetic excitations and free fermion behavior near the saturation field in \YbAl. This is surprising because it requires fine tuning of the anisotropic superexchange \cite{Wu2016} and magnetic dipole interaction between the dimer moments. In \YbAl\ the effective $g$-factor, which relates magnetic moment to the effective spin-1/2, $M^\alpha = \mu_B g_\alpha S^\alpha$, is very anisotropic, $g_z \gg g_{x,y} \simeq 0$ \cite{Wu2019,WuPRB2019}. Hence, magnetic dipole interaction, $\sim \bM_1 \bM_2$, provides anisotropic, $\sim S^z_1S^z_2$ contribution to the effective spin coupling.

To account for the contribution of the longer structures we also included the trimer subsystem, and now the total specific heat is described as:
\begin{eqnarray}
C_{\mathrm{Mag}}(T) = x^2(1-x)^2(C_{\mathrm{Dimer}}(T) +xC_{\mathrm{Trimer}}(T)).
\label{Eq:Specific_heat_2}
\end{eqnarray}
The energy splitting for the trimer subsystem at zero field was determined by solving Schrodinger equation for the linear spin trimer with isotropic exchange interaction: $J(\bS_1\bS_2 + \bS_2\bS_3)\psi = E\psi$ [the trimer Hamiltonian can be diagonalized as $\frac{1}{2} J \left( (\bS_1 + \bS_2 + \bS_3)^2 - (\bS_1 + \bS_3)^2 \right)$]. The splitting scheme is as follows: a doublet ground state, a doublet at $E_1 = J$, and a quartet at $E_2 = \frac{3}{2} J$ (note that due to the half-integer total spin of the trimer all states have Kramers degeneracy). The fitting of specific heat with Eq.~(\ref{Eq:Specific_heat_2}) revealed slightly lower Yb concentration, $x = 0.0430(2)$, whereas the energy splitting remains essentially unchanged, $\Delta_{\pm} = 0.212(5)$~meV, $\Delta_{0} = 0.211(5)$~meV.

Specific heat of \YYbAl\ at $B = 0$ calculated for $x = 0.043$ is shown in Fig.~\ref{SpecificHeat} (b). The blue and red lines show only the dimer and the dimer plus trimer contributions, respectively. Good agreement between the observed and calculated curves proves the predominantly dimer origin of the observed signal and gives an estimate of the contribution from the longer structures.

Having determined the real concentration of Yb in the sample, let us turn to the analysis of the specific heat measured in magnetic field. Application of magnetic field modifies the dimer spectrum according to Eqs.~\eqref{Zeeman_splitting_lowB},\eqref{Zeeman_splitting_highB} and splits the degenerate doublet state of Yb monomers according to $E(B) = g_a\mu_{\mathrm{B}}B$ [Fig.~\ref{Fig1:dimer_levels} (c,d)]. Thereby, we fit the measured specific heat as a sum of Yb monomer and dimer contributions~\footnote{Trimer contribution is below 1~\% of the in-filed specific heat, which is beyond the precision of our measurements.}:
\begin{eqnarray}
C_{\mathrm{Mag}}(T) = x(1-x)^2(C_{\mathrm{Mono}}(T) +xC_{\mathrm{Dimer}}(T)).
\label{Eq:Specific_heat_3}
\end{eqnarray}
In the fit, we fixed $x = 0.043$, while the doublet splitting, $E_{{i}}$, which determines the $g$ factor for Yb$_1$ subsystem were varied as free parameters. The experimental data along with the fitted curves are shown in Fig.~\ref{SpecificHeat} (a) by symbols and solid lines, respectively, and one can see that they are in good agreement. The extracted $g$-factor that describes the Zeeman splitting of Yb$_1$ [Fig.~\ref{SpecificHeat} (c)] is $g_a = 6.21(3)$, somewhat below $g_a = 6.9$ observed in pure \YbAl\ \cite{Wu2019}.

\subsection{Magnetization}

\begin{figure}[tb!]	
\center{\includegraphics[width=1\linewidth]{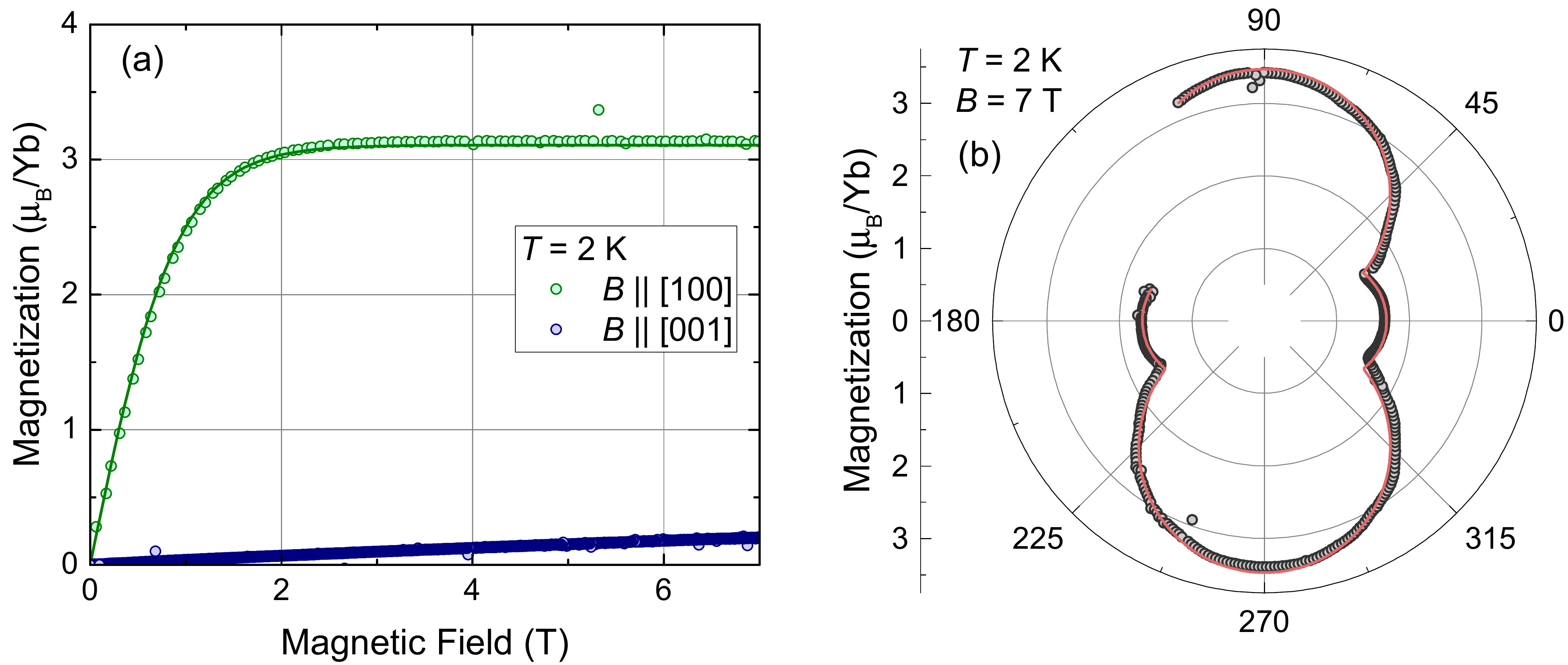}\vspace{-10pt}}
\caption{~(a) Field dependence of magnetization measured at $T = 2$~K along [100] and [001] directions. Green solid line represents the fit of $M_a(B)$ with Brillouin function.
 		 (b)~Angular dependence of magnetization in $ab$-plane measured at $B = 7$~T, $T = 2$~K. 0$^{\circ}$ and 90$^{\circ}$ corresponds to $B\parallel b$ and $B\parallel a$, respectively. Red line shows the fit with Eq.~(\ref{MvsA}).
}
\label{Magnetization}\vspace{-12pt}
\end{figure}

The local environment of Yb ions in Yb:YAlO$_3$ is very similar to the parent compound \YbAl, thus one can expect that the crystalline electric field produces a similar type of magnetic anisotropy in both compounds.
To verify this, we measured magnetization of \YYbAl\ at $T = 2$~K using MPMS3-VSM with horizontal rotator and the results are summarized in Fig.~\ref{Magnetization}.

In orthorhombic perovskite structure, Yb moments can either point along the $c$-axis, or lie in $ab$-plane. As the first step, we measured the field dependence of magnetization with $B\parallel [001]$ and $B\parallel [100]$ and the results are shown in Fig.~\ref{Magnetization}~(a).
One can see that the signal for $B$ along $[001]$ direction is very small ($\sim0.2$~$\mu_{\mathrm{B}}$ at 7~T) compared to the $B\parallel [100]$ case, indicating that the moments are indeed confined in the $ab$-plane in a similar fashion to that observed in other Yb-based orthorhombic perovskites~\cite{Wu2019, WuPRB2019, Nikitin2018}.
{By fitting $M_a(B)$ and $M_c(B)$ curves with the Brillouin function for $S = 1/2$, we found $g_a = 6.23$, in good agreement with the heat capacity results, and $g_c = g_\perp = 0.6$.}

To further investigate magnetic anisotropy within the $ab$ plane, we measured the angular dependence of magnetization using the horizontal rotator, as shown in Fig.~\ref{Magnetization}~(b). Note that at high magnetic field, above the saturation, Yb moments are confined in $ab$-plane by CEF and the angular dependence of magnetization can be described as~\cite{Wu2017, Wu2019},
\begin{align}
M(\theta) \approx \frac{M_{\mathrm{S}}}{2}\left(\lvert\cos\left(\theta-\varphi\right)\rvert + \lvert\cos\left(\theta+\varphi\right)\rvert\right) ,
\label{MvsA}
\end{align}
where $M_{\mathrm{S}}$ is the saturation moment of Yb, $\theta$ is the angle between the applied field and the $a$-axis, and $\varphi$ is the angle between the $a$-axis and the Yb$^{3+}$ moment direction. Result of the fit is shown in Fig.~\ref{Magnetization} by red line. Our analysis revealed $M_{\mathrm{S}} = 3.57~\mu_{\mathrm{B}}$/Yb and $\varphi = 26^{\circ}$, which are close to to $M_{\mathrm{S}} = 3.8~\mu_{\mathrm{B}}$/Yb and $\varphi = 23.5^{\circ}$ observed in pure \YbAl. In agreement with the above heat capacity results, the $g$-factor along the quantization axis is $g_z = 7.14$ ($g_a = 6.4$), somewhat lower than $g_z = 7.6$ found in \YbAl\ \cite{WuPRB2019}. $M(B)$ and $M(\theta)$ curves shown in Fig.~\ref{Magnetization} were measured on different small single crystals with $m \sim 8$~mg; a slight discrepancy of $M_{\mathrm{S}}$ and $g_a$ obtained from these measurements reflects an uncertainty of sample weight.

\subsection{Neutron scattering}

\begin{figure*}[tb!]	
\center{\includegraphics[width=.9\linewidth]{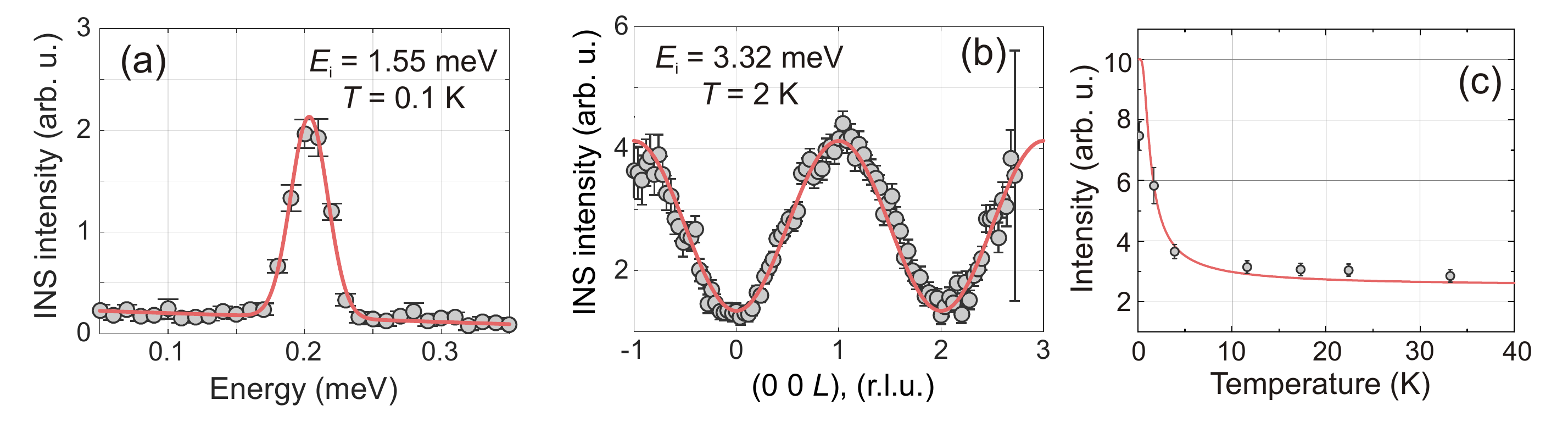}\vspace{-10pt}}
\caption{~Inelastic neutron scattering spectra of \YYbAl.
(a)~Energy spectrum measured at $T \approx 0.1$~K, $E_{\mathrm{i}} = 1.55$~meV. The data were integrated within $H = 0\pm0.1$, $K = 1 \pm 0.5$, $L = 1 \pm 0.5$.
(b)~Constant-energy cut along the $(00L)$ direction taken at $T \approx 2$~K, $E_{\mathrm{i}} = 3.32$~meV. The data were integrated in energy window $\hbar\omega = [0.17$--$0.25]$~meV; $H = 0\pm0.1$, $K = 0\pm1$.
(c)~Intensity of the inelastic peak at $\mathbf{Q} = (001)$ as a function of temperature and its fit with Boltzmann function.
Error bars represent one standard deviation.
}
\label{Neutron_cuts}\vspace{-12pt}	
\end{figure*}

\begin{figure}[b!]	
\center{\includegraphics[width=1\linewidth]{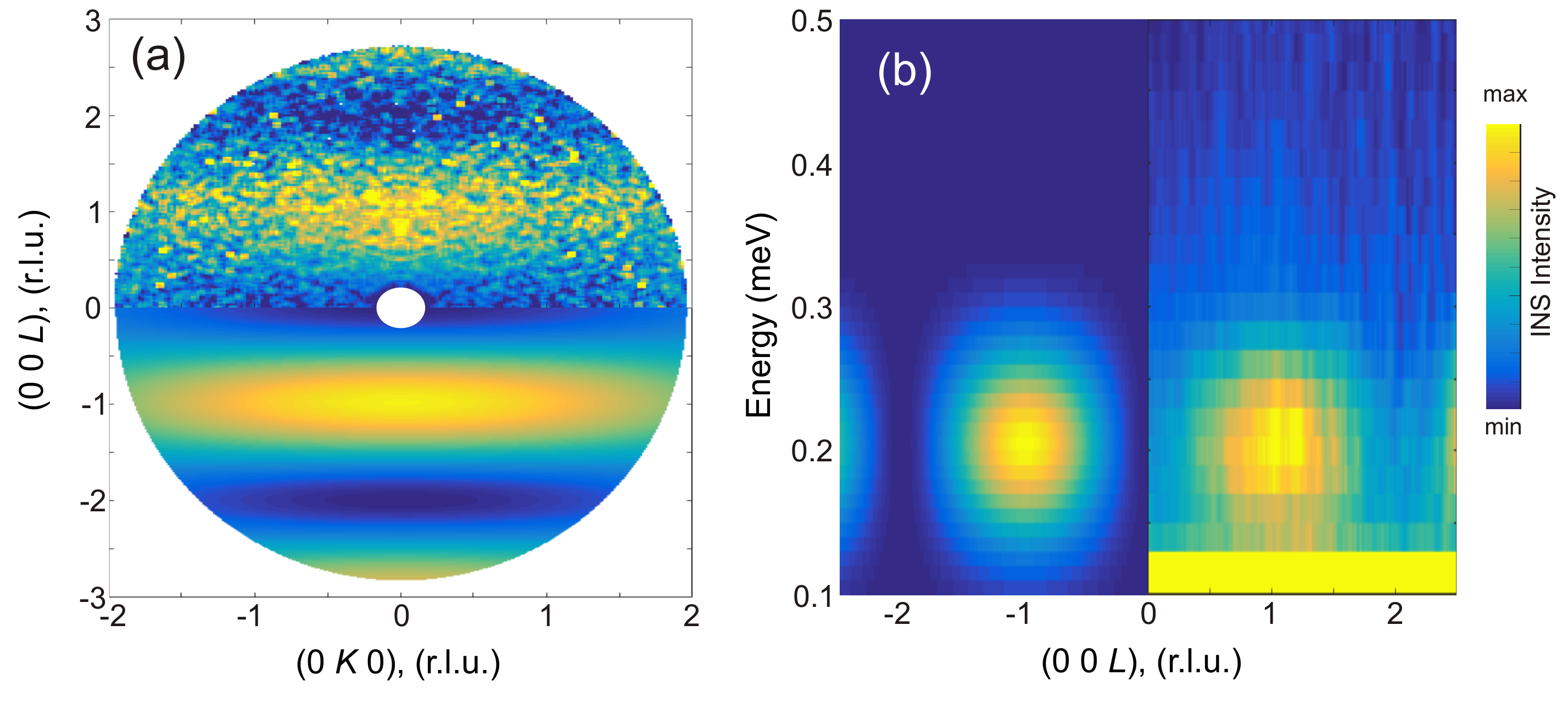}\vspace{-10pt}}
\caption{~Inelastic neutron scattering spectra of \YYbAl.
 		     (a)~Observed (top) and modeled (bottom) constant energy slice in $(0KL)$ scattering plane at $T \approx 2$~K, $E_{\mathrm{i}} = 3.32$~meV. The data are integrated within the energy window $\hbar\omega = [0.17, 0.25]$~meV; $H = 0\pm0.1$.
 		     (b)~Observed (right) and modeled (left) spectra along the $(00L)$ direction at $T \approx 2$~K, $E_{\mathrm{i}} = 3.32$~meV; $K = 0\pm1$;  $H = 0\pm0.1$.
}
\label{Neutron_slice}
\end{figure}

Inelastic neutron scattering (INS) is a powerful tool for investigation of magnetic excitations (see Ref.~\cite{Squires_book_1978,lovesey1984theory,JensenMackintosh_book1991,ZaliznyakLee_MNSChapter} and references therein) and has proven extremely successful in characterizing magnetic clusters \cite{furrer2013magnetic,Kenzelmann_PRL2003}. We write the INS cross-section in the following form~\cite{Squires_book_1978,JensenMackintosh_book1991,ZaliznyakLee_MNSChapter,lovesey1984theory},
\begin{eqnarray}
\frac{d^2\sigma}{d\Omega\omega} \propto &|F(\mathbf{Q})|^2 \sum_{\alpha,\beta} \frac{g_\alpha}{2} \frac{g_\beta}{2} \left(\delta_{\alpha \beta} -\frac{Q_\alpha Q_\beta}{Q^2} \right) \sum_{{i,j}} e^{i\mathbf{Q \cdot r_{\textit{ij}}}} \nonumber \\
\times \sum_{\lambda, \lambda'} & p_\lambda \langle\lambda|\hat{S}_{i}^{\alpha}|\lambda'\rangle \langle\lambda'|\hat{S}_{j}^{\beta}|\lambda\rangle \delta(\hbar\omega + E_\lambda - E_{\lambda'}), \;\;\;\;\;\;\;\;
\label{INS_Cross_section}
\end{eqnarray}
corresponding to unpolarized neutrons used in our measurements. Here, $F(\mathbf{Q})$ is magnetic form factor of Yb$^{3+}$; $\alpha, \beta = x, y, z$ and $g_\alpha$, $g_\beta$ are the corresponding $g$-factors; $|\lambda\rangle$ and $|\lambda'\rangle$ are the eigenvectors of the initial and final state of the system; $p_\lambda$ is the probability to find the system in state $\lambda$ [cf Eq.~\eqref{Eq:T_prefactor}]; $\hat{S}_{i}^{\alpha}$ and $\hat{S}_{j}^{\beta}$ are the effective spin-1/2 angular momentum operators at sites $i$ and $j$, respectively.
The term $P = \sum_{\alpha,\beta}\big(\delta_{\alpha,\beta} -\frac{Q_\alpha Q_\beta}{Q^2} \big)$ is a so-called polarization factor of neutron scattering. It takes into account the property of magnetic dipole interaction that only magnetic moment components that are perpendicular to $\mathbf{Q}$ contribute to magnetic neutron scattering cross-section.
However, the polarization factor is irrelevant for the present study because our spectra were collected within $(0KL)$ scattering plane, whereas the Yb moments have a preferred orientation along the [100]-axis. Therefore, the dominant fluctuations that are favorably weighted by the $g$-factor in Eq.~\eqref{INS_Cross_section} have $P \simeq 1$ for our scattering plane.
$e^{i\mathbf{Q \cdot r_{\textit{ij}}}}$ gives rise to the structure factor, which causes the modulation of INS intensity in reciprocal space according to the geometry of the cluster. For Yb dimers in \YYbAl, $\mathbf{r}_{12} \approx (0, 0, \frac{1}{2})$~\footnote{In real structure $\mathbf{r}_{12} \approx (\delta_a, \delta_b, \frac{1}{2})$, with $\delta_a \approx \pm 0.03$ and $\delta_b \approx \pm 0.13$.
This causes an additional intensity modulation $I(\mathbf{Q}) \propto \text{cos}(2\pi\delta_a)\cdot\text{cos}(2\pi\delta_b)$. E.g. it is responsible for the second ``shadow mode'' in YbFeO$_3$ spectrum~\cite{Nikitin2018}. However, it has only a minor influence at our range of $\mathbf{Q}$.} and the structure factor can be reduced to $S(\mathbf{Q}) = 1-\text{cos}(Q_L\pi)$~\cite{haraldsen2005neutron}.

At low temperature, $k_{\mathrm{B}}T\ll J$, the spectrum of a dimer measured by INS should consist of a single peak at $E = \Delta_0 = J$, which corresponds to $\Delta S^z = 0$ transition. The peak at $\Delta_{\pm} = J \frac{1 + \Delta}{2}$ corresponding to $\Delta S^z = \pm 1$ is relatively suppressed by the $g$-factor ratio, $(g_c/g_{a,b})^2 \gtrsim 100$ \cite{WuPRB2019}, and is undetectable. In pure \YbAl, exchange interaction was found to be $J = 0.21$~meV~\cite{Wu2019, WuPRB2019}. To explore the dynamic response of \YYbAl, we performed high-resolution INS measurements. The INS spectrum shows a sharp, resolution-limited inelastic peak, which can be fitted by a Gaussian function with the center energy $E = 0.2034(4)$~meV [Fig.~\ref{Neutron_cuts}~(a)], very close to the energy of exchange interaction found in pure \YbAl.

To further elucidate the dimer origin of the observed peak, we studied the $\mathbf{Q}$-dependence of the excitation using the same instrument with orange cryostat (base temperature $\sim$2~K) and higher neutron incident energies in order to cover large area of $\mathbf{Q}$-space.
The $\mathbf{Q}$-dependence of the observed excitation is presented in Fig.~\ref{Neutron_slice}~(a), which shows the constant-energy map in $(0KL)$ plane integrated within the energy window of the excitation~\footnote{To increase the signal-to-noise ratio we symmetrized the data according to the crystal symmetry and then unfolded it back, which means that data at negative and positive $K$ in Fig.~\ref{Neutron_slice}~(a) are equivalent}.
One can see that the spectral intensity is modulated along the $L$-direction because of the structure factor and shows stripe-like features elongated along the $K$-direction.
Figure~\ref{Neutron_slice}~(b) presents the INS spectrum along $(00L)$ direction where one can see the gapped excitation at $E \approx 0.2$~meV whose intensity is concentrated close to (001).

The INS intensity along the $(00L)$ direction integrated within the $\hbar\omega = [0.17, 0.25]$~meV energy range is shown in Fig.~\ref{Neutron_cuts}~(b). We fitted the obtained curve using the equation for the dimer's structure factor plus a constant background, $I(Q_L) = a_0 + b(1-\text{cos}(Q_L\pi))$, and found a very good agreement between the experimental spectrum and the fitted curve. The temperature dependence of the intensity of the inelastic peak at $\mathbf{Q} = (001)$ fits well with the thermal population factor of the ground state, Eq.~\eqref{Eq:T_prefactor} [Fig.~\ref{Neutron_cuts}~(c)]. The gap, $\Delta = 0.21(5)$~meV, obtained from this fit agrees well with that determined directly.

We modeled the experimental spectrum and constant-energy map taking into account Yb$^{3+}$ form factor and dimer structure factor, experimental energy resolution, and the observed transition energy $E = 0.2$~meV. The results of the modeling show a nearly perfect agreement between the experimental and calculated patterns, which provides another evidence of the dimer origin of the observed excitation [Figs.~\ref{Neutron_slice}~(a-b)].

\section{Discussion and Conclusion}

To summarize, we performed detailed experimental investigation of Yb-Yb magnetic interactions in diluted \YYbAl, which sheds the light on magnetic Hamiltonian of Yb dimers, n-mers, and pure \YbAl. For Yb-Yb dimers, we determined the level splitting and the effective spin-1/2 coupling, $J$, from the heat capacity and high-resolution neutron scattering measurements. From the magnetic field dependent heat capacity and magnetization, we determined the concentration of the isolated Yb$^{3+}$ moments and their effective $g$-factor.

To corroborate that the measurement of substituted \YYbAl\ indeed reflects physics of \YbAl, we want to stress several key points of our analysis. First of all, the saturation magnetization per Yb ion and the magnetic anisotropy are very similar for both samples, which means that the CEF and the ground state doublet wavefunctions are also virtually identical. Second, the inelastic peak in the INS spectrum of \YYbAl\ is located at $E = 0.2$~meV, very close to the value of exchange interaction determined by fitting the spectrum of magnetic excitations in \YbAl~\cite{Wu2019}. The structure factor of the observed excitation exhibits a simple cosine modulation, as expected for magnetic dimer. Third, we measured magnetic specific heat with and without mangetic field. In zero magnetic field the dominant contribution is from Yb-Yb dimers and determines the dimer level splitting. In magnetic field, it is dominated by monomers (single Yb moments) and can be quantitatively described using a single concentration parameter, $x$, which corroborates random statistical distribution of the Yb ions in the sample. In addition to $x$, we have also refined the $g$-factor describing Zeeman splitting of a single Yb$^{3+}$ doublet in \YYbAl\ in magnetic field and obtained the value that is within 10\% of that determined in \YbAl.

We would like to point out that in-field low-temperature specific heat measurements, such as described in Sec.~\ref{Sec:SpecHeat}, provide an extremely sensitive quantitative experimental method for determining the concentration of magnetic impurities in a bulk sample. Its sensitivity and capability of averaging over the large sample volume are well beyond the standard non-destructive analytical techniques, such as energy dispersive x-ray spectroscopy analysis.
An alternative bulk-sensitive thermodynamic probe is the magnetization measurement. Magnetic impurities exhibit paramagnetic behavior, which is well understood and can be described by Brillouin function, allowing to extract the total number of magnetic ions in the sample. However, such an analysis requires \textit{a-priori} knowledge of the magnetic moment of the impurity ion in the given CEF environment. This might not be known, especially for the $4f$ ions with strong spin-orbit coupling, where CEF determines the magnetic moment. The specific heat measurements have no such limitation and we suggest this method for measurements of impurity concentration for samples with $4f$ magnetic ions, especially for the cases where concentration of magnetic ions is relatively low.

Finally and perhaps most surprisingly, the analysis of the zero-field magnetic specific heat have shown that there is no noticeable splitting of the excited triplet state, which is further corroborated by comparing the obtained energy of the triplet level with the high-resolution INS result. Hence, we conclude that the dominant, $c$-axis exchange interaction of the effective spins-1/2 is surprisingly isotropic despite the low site symmetry of Yb ions, in agreement with what was inferred in the previous report on parent compound \YbAl~\cite{Wu2019}. As discussed above, this is surprising because a non-negligible magnetic dipole interaction of the nearest Yb moments, $\frac{\mu_0}{4\pi (c/2)^3} \bM_1 \bM_2$ ($\mu_0$ is the magnetic constant), yields a measurable anisotropic contribution to the effective spin interaction, $\sim (10^{-3}$~meV$)\cdot \sum_\alpha g_\alpha^2  S^\alpha_1S^\alpha_2$, which appears to be balanced by the anisotropic exchange coupling, suggesting serendipitous fine-tuning. Our results provide important basis for understanding quantum states of Yb n-mers and chains in Yb:YAlO$_3$ system and for their possible future applications in quantum information science.

\section*{Acknowledgments}
We acknowledge A. S. Sukhanov for stimulating discussions.
This research used resources at the Spallation Neutron Source, a DOE Office of Science User Facility operated by Oak Ridge National Laboratory.
S.E.N. acknowledges support from the International Max Planck Research School for Chemistry and Physics of Quantum Materials (IMPRS-CPQM).
Laue x-ray diffraction measurements were conducted at the Center for Nanophase Materials Sciences (CNMS) (CNMS2019-R18) at Oak Ridge National Laboratory (ORNL), which is a DOE Office of Science User Facility. The work at Brookhaven National Laboratory was supported by Office of Basic Energy Sciences (BES), Division of Materials Sciences and Engineering,  U.S. Department of Energy (DOE), under contract DE-SC0012704.


\begin{thebibliography}{49}%
\makeatletter
\providecommand \@ifxundefined [1]{%
 \@ifx{#1\undefined}
}%
\providecommand \@ifnum [1]{%
 \ifnum #1\expandafter \@firstoftwo
 \else \expandafter \@secondoftwo
 \fi
}%
\providecommand \@ifx [1]{%
 \ifx #1\expandafter \@firstoftwo
 \else \expandafter \@secondoftwo
 \fi
}%
\providecommand \natexlab [1]{#1}%
\providecommand \enquote  [1]{``#1''}%
\providecommand \bibnamefont  [1]{#1}%
\providecommand \bibfnamefont [1]{#1}%
\providecommand \citenamefont [1]{#1}%
\providecommand \href@noop [0]{\@secondoftwo}%
\providecommand \href [0]{\begingroup \@sanitize@url \@href}%
\providecommand \@href[1]{\@@startlink{#1}\@@href}%
\providecommand \@@href[1]{\endgroup#1\@@endlink}%
\providecommand \@sanitize@url [0]{\catcode `\\12\catcode `\$12\catcode
  `\&12\catcode `\#12\catcode `\^12\catcode `\_12\catcode `\%12\relax}%
\providecommand \@@startlink[1]{}%
\providecommand \@@endlink[0]{}%
\providecommand \url  [0]{\begingroup\@sanitize@url \@url }%
\providecommand \@url [1]{\endgroup\@href {#1}{\urlprefix }}%
\providecommand \urlprefix  [0]{URL }%
\providecommand \Eprint [0]{\href }%
\providecommand \doibase [0]{http://dx.doi.org/}%
\providecommand \selectlanguage [0]{\@gobble}%
\providecommand \bibinfo  [0]{\@secondoftwo}%
\providecommand \bibfield  [0]{\@secondoftwo}%
\providecommand \translation [1]{[#1]}%
\providecommand \BibitemOpen [0]{}%
\providecommand \bibitemStop [0]{}%
\providecommand \bibitemNoStop [0]{.\EOS\space}%
\providecommand \EOS [0]{\spacefactor3000\relax}%
\providecommand \BibitemShut  [1]{\csname bibitem#1\endcsname}%
\let\auto@bib@innerbib\@empty
\bibitem [{\citenamefont {Wu}\ \emph {et~al.}(2019{\natexlab{a}})\citenamefont
  {Wu}, \citenamefont {Nikitin}, \citenamefont {Wang}, \citenamefont {Zhu},
  \citenamefont {Batista}, \citenamefont {Tsvelik}, \citenamefont {Samarakoon},
  \citenamefont {Tennant}, \citenamefont {Brando}, \citenamefont {Vasylechko},
  \citenamefont {Frontzek}, \citenamefont {Savici}, \citenamefont {Sala},
  \citenamefont {Ehlers}, \citenamefont {Christianson}, \citenamefont
  {Lumsden},\ and\ \citenamefont {Podlesnyak}}]{Wu2019}%
  \BibitemOpen
  \bibfield  {author} {\bibinfo {author} {\bibfnamefont {L.~S.}\ \bibnamefont
  {Wu}}, \bibinfo {author} {\bibfnamefont {S.~E.}\ \bibnamefont {Nikitin}},
  \bibinfo {author} {\bibfnamefont {Z.}~\bibnamefont {Wang}}, \bibinfo {author}
  {\bibfnamefont {W.}~\bibnamefont {Zhu}}, \bibinfo {author} {\bibfnamefont
  {C.~D.}\ \bibnamefont {Batista}}, \bibinfo {author} {\bibfnamefont {A.~M.}\
  \bibnamefont {Tsvelik}}, \bibinfo {author} {\bibfnamefont {A.~M.}\
  \bibnamefont {Samarakoon}}, \bibinfo {author} {\bibfnamefont {D.~A.}\
  \bibnamefont {Tennant}}, \bibinfo {author} {\bibfnamefont {M.}~\bibnamefont
  {Brando}}, \bibinfo {author} {\bibfnamefont {L.}~\bibnamefont {Vasylechko}},
  \bibinfo {author} {\bibfnamefont {M.}~\bibnamefont {Frontzek}}, \bibinfo
  {author} {\bibfnamefont {A.~T.}\ \bibnamefont {Savici}}, \bibinfo {author}
  {\bibfnamefont {G.}~\bibnamefont {Sala}}, \bibinfo {author} {\bibfnamefont
  {G.}~\bibnamefont {Ehlers}}, \bibinfo {author} {\bibfnamefont {A.~D.}\
  \bibnamefont {Christianson}}, \bibinfo {author} {\bibfnamefont {M.~D.}\
  \bibnamefont {Lumsden}}, \ and\ \bibinfo {author} {\bibfnamefont
  {A.}~\bibnamefont {Podlesnyak}},\ }\bibfield  {title} {\enquote {\bibinfo
  {title} {{Tomonaga-Luttinger liquid behavior and spinon confinement in
  YbAlO$_3$}},}\ }\href {\doibase 10.1038/s41467-019-08485-7} {\bibfield
  {journal} {\bibinfo  {journal} {Nat. Commun.}\ }\textbf {\bibinfo {volume}
  {10}},\ \bibinfo {pages} {698} (\bibinfo {year}
  {2019}{\natexlab{a}})}\BibitemShut {NoStop}%
\bibitem [{\citenamefont {Wu}\ \emph {et~al.}(2019{\natexlab{b}})\citenamefont
  {Wu}, \citenamefont {Nikitin}, \citenamefont {Brando}, \citenamefont
  {Vasylechko}, \citenamefont {Ehlers}, \citenamefont {Frontzek}, \citenamefont
  {Savici}, \citenamefont {Sala}, \citenamefont {Christianson}, \citenamefont
  {Lumsden},\ and\ \citenamefont {Podlesnyak}}]{WuPRB2019}%
  \BibitemOpen
  \bibfield  {author} {\bibinfo {author} {\bibfnamefont {L.~S.}\ \bibnamefont
  {Wu}}, \bibinfo {author} {\bibfnamefont {S.~E.}\ \bibnamefont {Nikitin}},
  \bibinfo {author} {\bibfnamefont {M.}~\bibnamefont {Brando}}, \bibinfo
  {author} {\bibfnamefont {L.}~\bibnamefont {Vasylechko}}, \bibinfo {author}
  {\bibfnamefont {G.}~\bibnamefont {Ehlers}}, \bibinfo {author} {\bibfnamefont
  {M.}~\bibnamefont {Frontzek}}, \bibinfo {author} {\bibfnamefont {A.~T.}\
  \bibnamefont {Savici}}, \bibinfo {author} {\bibfnamefont {G.}~\bibnamefont
  {Sala}}, \bibinfo {author} {\bibfnamefont {A.~D.}\ \bibnamefont
  {Christianson}}, \bibinfo {author} {\bibfnamefont {M.~D.}\ \bibnamefont
  {Lumsden}}, \ and\ \bibinfo {author} {\bibfnamefont {A.}~\bibnamefont
  {Podlesnyak}},\ }\bibfield  {title} {\enquote {\bibinfo {title}
  {{Antiferromagnetic ordering and dipolar interactions of
  ${\mathrm{YbAlO}}_{3}$}},}\ }\href {\doibase 10.1103/PhysRevB.99.195117}
  {\bibfield  {journal} {\bibinfo  {journal} {Phys. Rev. B}\ }\textbf {\bibinfo
  {volume} {99}},\ \bibinfo {pages} {195117} (\bibinfo {year}
  {2019}{\natexlab{b}})}\BibitemShut {NoStop}%
\bibitem [{\citenamefont {Wu}\ \emph {et~al.}(2016)\citenamefont {Wu},
  \citenamefont {Gannon}, \citenamefont {Zaliznyak}, \citenamefont {Tsvelik},
  \citenamefont {Brockmann}, \citenamefont {Caux}, \citenamefont {Kim},
  \citenamefont {Qiu}, \citenamefont {Copley}, \citenamefont {Ehlers},
  \citenamefont {Podlesnyak},\ and\ \citenamefont {Aronson}}]{Wu2016}%
  \BibitemOpen
  \bibfield  {author} {\bibinfo {author} {\bibfnamefont {L.~S.}\ \bibnamefont
  {Wu}}, \bibinfo {author} {\bibfnamefont {W.~J.}\ \bibnamefont {Gannon}},
  \bibinfo {author} {\bibfnamefont {I.~A.}\ \bibnamefont {Zaliznyak}}, \bibinfo
  {author} {\bibfnamefont {A.~M.}\ \bibnamefont {Tsvelik}}, \bibinfo {author}
  {\bibfnamefont {M.}~\bibnamefont {Brockmann}}, \bibinfo {author}
  {\bibfnamefont {J.-S.}\ \bibnamefont {Caux}}, \bibinfo {author}
  {\bibfnamefont {M.~S.}\ \bibnamefont {Kim}}, \bibinfo {author} {\bibfnamefont
  {Y.}~\bibnamefont {Qiu}}, \bibinfo {author} {\bibfnamefont {J.~R.~D.}\
  \bibnamefont {Copley}}, \bibinfo {author} {\bibfnamefont {G.}~\bibnamefont
  {Ehlers}}, \bibinfo {author} {\bibfnamefont {A.}~\bibnamefont {Podlesnyak}},
  \ and\ \bibinfo {author} {\bibfnamefont {M.~C.}\ \bibnamefont {Aronson}},\
  }\bibfield  {title} {\enquote {\bibinfo {title} {{Orbital-exchange and
  fractional quantum number excitations in an f-electron metal,
  Yb$_2$Pt$_2$Pb}},}\ }\href {\doibase 10.1126/science.aaf0981} {\bibfield
  {journal} {\bibinfo  {journal} {Science}\ }\textbf {\bibinfo {volume}
  {352}},\ \bibinfo {pages} {1206} (\bibinfo {year} {2016})}\BibitemShut
  {NoStop}%
\bibitem [{\citenamefont {Gannon}\ \emph {et~al.}(2019)\citenamefont {Gannon},
  \citenamefont {Zaliznyak}, \citenamefont {Wu}, \citenamefont {Feiguin},
  \citenamefont {Tsvelik}, \citenamefont {Demmel}, \citenamefont {Qiu},
  \citenamefont {Copley}, \citenamefont {Kim},\ and\ \citenamefont
  {Aronson}}]{Gannon2019}%
  \BibitemOpen
  \bibfield  {author} {\bibinfo {author} {\bibfnamefont {W.~J.}\ \bibnamefont
  {Gannon}}, \bibinfo {author} {\bibfnamefont {I.~A.}\ \bibnamefont
  {Zaliznyak}}, \bibinfo {author} {\bibfnamefont {L.~S.}\ \bibnamefont {Wu}},
  \bibinfo {author} {\bibfnamefont {A.~E.}\ \bibnamefont {Feiguin}}, \bibinfo
  {author} {\bibfnamefont {A.~M.}\ \bibnamefont {Tsvelik}}, \bibinfo {author}
  {\bibfnamefont {F.}~\bibnamefont {Demmel}}, \bibinfo {author} {\bibfnamefont
  {Y.}~\bibnamefont {Qiu}}, \bibinfo {author} {\bibfnamefont {J.~R.~D.}\
  \bibnamefont {Copley}}, \bibinfo {author} {\bibfnamefont {M.~S.}\
  \bibnamefont {Kim}}, \ and\ \bibinfo {author} {\bibfnamefont {M.~C.}\
  \bibnamefont {Aronson}},\ }\bibfield  {title} {\enquote {\bibinfo {title}
  {{Spinon confinement and a sharp longitudinal mode in Yb$_2$Pt$_2$Pb in
  magnetic fields}},}\ }\href {\doibase 10.1038/s41467-019-08715-y} {\bibfield
  {journal} {\bibinfo  {journal} {Nat. Commun.}\ }\textbf {\bibinfo {volume}
  {10}},\ \bibinfo {pages} {1123} (\bibinfo {year} {2019})}\BibitemShut
  {NoStop}%
\bibitem [{\citenamefont {Bertaina}\ \emph {et~al.}(2007)\citenamefont
  {Bertaina}, \citenamefont {Gambarelli}, \citenamefont {Tkachuk},
  \citenamefont {Kurkin}, \citenamefont {Malkin}, \citenamefont {Stepanov},\
  and\ \citenamefont {Barbara}}]{Bertaina2007}%
  \BibitemOpen
  \bibfield  {author} {\bibinfo {author} {\bibfnamefont {S.}~\bibnamefont
  {Bertaina}}, \bibinfo {author} {\bibfnamefont {S.}~\bibnamefont
  {Gambarelli}}, \bibinfo {author} {\bibfnamefont {A.}~\bibnamefont {Tkachuk}},
  \bibinfo {author} {\bibfnamefont {I.~N.}\ \bibnamefont {Kurkin}}, \bibinfo
  {author} {\bibfnamefont {B.}~\bibnamefont {Malkin}}, \bibinfo {author}
  {\bibfnamefont {A.}~\bibnamefont {Stepanov}}, \ and\ \bibinfo {author}
  {\bibfnamefont {B.}~\bibnamefont {Barbara}},\ }\bibfield  {title} {\enquote
  {\bibinfo {title} {Rare-earth solid-state qubits},}\ }\href {\doibase
  10.1038/nnano.2006.174} {\bibfield  {journal} {\bibinfo  {journal} {Nat.
  Nanotechnol.}\ }\textbf {\bibinfo {volume} {2}},\ \bibinfo {pages} {39--42}
  (\bibinfo {year} {2007})}\BibitemShut {NoStop}%
\bibitem [{\citenamefont {Simon}\ \emph {et~al.}(2010)\citenamefont {Simon},
  \citenamefont {Afzelius}, \citenamefont {Appel}, \citenamefont {Boyer de~la
  Giroday}, \citenamefont {Dewhurst}, \citenamefont {Gisin}, \citenamefont
  {Hu}, \citenamefont {Jelezko}, \citenamefont {Kr\"{o}ll}, \citenamefont
  {M\"{u}ller}, \citenamefont {Nunn}, \citenamefont {Polzik}, \citenamefont
  {Rarity}, \citenamefont {De~Riedmatten}, \citenamefont {Rosenfeld},
  \citenamefont {Shields}, \citenamefont {Sk\"{o}ld}, \citenamefont
  {Stevenson}, \citenamefont {Thew}, \citenamefont {Walmsley}, \citenamefont
  {Weber}, \citenamefont {Weinfurter}, \citenamefont {Wrachtrup},\ and\
  \citenamefont {Young}}]{Simon2010}%
  \BibitemOpen
  \bibfield  {author} {\bibinfo {author} {\bibfnamefont {C.}~\bibnamefont
  {Simon}}, \bibinfo {author} {\bibfnamefont {M.}~\bibnamefont {Afzelius}},
  \bibinfo {author} {\bibfnamefont {J.}~\bibnamefont {Appel}}, \bibinfo
  {author} {\bibfnamefont {A.}~\bibnamefont {Boyer de~la Giroday}}, \bibinfo
  {author} {\bibfnamefont {S.~J.}\ \bibnamefont {Dewhurst}}, \bibinfo {author}
  {\bibfnamefont {N.}~\bibnamefont {Gisin}}, \bibinfo {author} {\bibfnamefont
  {C.~Y.}\ \bibnamefont {Hu}}, \bibinfo {author} {\bibfnamefont
  {F.}~\bibnamefont {Jelezko}}, \bibinfo {author} {\bibfnamefont
  {S.}~\bibnamefont {Kr\"{o}ll}}, \bibinfo {author} {\bibfnamefont {J.~H.}\
  \bibnamefont {M\"{u}ller}}, \bibinfo {author} {\bibfnamefont
  {J.}~\bibnamefont {Nunn}}, \bibinfo {author} {\bibfnamefont {E.~S.}\
  \bibnamefont {Polzik}}, \bibinfo {author} {\bibfnamefont {J.~G.}\
  \bibnamefont {Rarity}}, \bibinfo {author} {\bibfnamefont {H.}~\bibnamefont
  {De~Riedmatten}}, \bibinfo {author} {\bibfnamefont {W.}~\bibnamefont
  {Rosenfeld}}, \bibinfo {author} {\bibfnamefont {A.~J.}\ \bibnamefont
  {Shields}}, \bibinfo {author} {\bibfnamefont {N.}~\bibnamefont {Sk\"{o}ld}},
  \bibinfo {author} {\bibfnamefont {R.~M.}\ \bibnamefont {Stevenson}}, \bibinfo
  {author} {\bibfnamefont {R.}~\bibnamefont {Thew}}, \bibinfo {author}
  {\bibfnamefont {I.~A.}\ \bibnamefont {Walmsley}}, \bibinfo {author}
  {\bibfnamefont {M.~C.}\ \bibnamefont {Weber}}, \bibinfo {author}
  {\bibfnamefont {H.}~\bibnamefont {Weinfurter}}, \bibinfo {author}
  {\bibfnamefont {J.}~\bibnamefont {Wrachtrup}}, \ and\ \bibinfo {author}
  {\bibfnamefont {R.~J.}\ \bibnamefont {Young}},\ }\bibfield  {title} {\enquote
  {\bibinfo {title} {Quantum memories},}\ }\href {\doibase
  10.1140/epjd/e2010-00103-y} {\bibfield  {journal} {\bibinfo  {journal} {Eur.
  Phys. J. D}\ }\textbf {\bibinfo {volume} {58}},\ \bibinfo {pages} {1--22}
  (\bibinfo {year} {2010})}\BibitemShut {NoStop}%
\bibitem [{\citenamefont {Thiel}\ \emph {et~al.}(2011)\citenamefont {Thiel},
  \citenamefont {B\"{o}ttger},\ and\ \citenamefont {Cone}}]{Thiel2011}%
  \BibitemOpen
  \bibfield  {author} {\bibinfo {author} {\bibfnamefont {C.W.}\ \bibnamefont
  {Thiel}}, \bibinfo {author} {\bibfnamefont {Thomas}\ \bibnamefont
  {B\"{o}ttger}}, \ and\ \bibinfo {author} {\bibfnamefont {R.L.}\ \bibnamefont
  {Cone}},\ }\bibfield  {title} {\enquote {\bibinfo {title} {Rare-earth-doped
  materials for applications in quantum information storage and signal
  processing},}\ }\href {\doibase https://doi.org/10.1016/j.jlumin.2010.12.015}
  {\bibfield  {journal} {\bibinfo  {journal} {J. Lumin.}\ }\textbf {\bibinfo
  {volume} {131}},\ \bibinfo {pages} {353 -- 361} (\bibinfo {year}
  {2011})}\BibitemShut {NoStop}%
\bibitem [{\citenamefont {Awschalom}\ \emph {et~al.}(2018)\citenamefont
  {Awschalom}, \citenamefont {Hanson}, \citenamefont {Wrachtrup},\ and\
  \citenamefont {Zhou}}]{Awschalom2018}%
  \BibitemOpen
  \bibfield  {author} {\bibinfo {author} {\bibfnamefont {D.~D.}\ \bibnamefont
  {Awschalom}}, \bibinfo {author} {\bibfnamefont {R.}~\bibnamefont {Hanson}},
  \bibinfo {author} {\bibfnamefont {J.}~\bibnamefont {Wrachtrup}}, \ and\
  \bibinfo {author} {\bibfnamefont {B.~B.}\ \bibnamefont {Zhou}},\ }\bibfield
  {title} {\enquote {\bibinfo {title} {Quantum technologies with optically
  interfaced solid-state spins},}\ }\href {\doibase 10.1038/s41566-018-0232-21}
  {\bibfield  {journal} {\bibinfo  {journal} {Nat. Photonics}\ }\textbf
  {\bibinfo {volume} {12}},\ \bibinfo {pages} {516--527} (\bibinfo {year}
  {2018})}\BibitemShut {NoStop}%
\bibitem [{\citenamefont {Kunkel}\ and\ \citenamefont
  {Goldner}(2018)}]{Kunkel2018}%
  \BibitemOpen
  \bibfield  {author} {\bibinfo {author} {\bibfnamefont {N.}~\bibnamefont
  {Kunkel}}\ and\ \bibinfo {author} {\bibfnamefont {P.}~\bibnamefont
  {Goldner}},\ }\bibfield  {title} {\enquote {\bibinfo {title} {{Recent
  Advances in Rare Earth Doped Inorganic Crystalline Materials for Quantum
  Information Processing}},}\ }\href {\doibase 10.1002/zaac.201700425}
  {\bibfield  {journal} {\bibinfo  {journal} {Z. Anorg. Allg. Chem}\ }\textbf
  {\bibinfo {volume} {644}},\ \bibinfo {pages} {66--76} (\bibinfo {year}
  {2018})}\BibitemShut {NoStop}%
\bibitem [{\citenamefont {Olmschenk}\ \emph {et~al.}(2009)\citenamefont
  {Olmschenk}, \citenamefont {Matsukevich}, \citenamefont {Maunz},
  \citenamefont {Hayes}, \citenamefont {Duan},\ and\ \citenamefont
  {Monroe}}]{Olmschenk2009}%
  \BibitemOpen
  \bibfield  {author} {\bibinfo {author} {\bibfnamefont {S.}~\bibnamefont
  {Olmschenk}}, \bibinfo {author} {\bibfnamefont {D.~N.}\ \bibnamefont
  {Matsukevich}}, \bibinfo {author} {\bibfnamefont {P.}~\bibnamefont {Maunz}},
  \bibinfo {author} {\bibfnamefont {D.}~\bibnamefont {Hayes}}, \bibinfo
  {author} {\bibfnamefont {L.-M.}\ \bibnamefont {Duan}}, \ and\ \bibinfo
  {author} {\bibfnamefont {C.}~\bibnamefont {Monroe}},\ }\bibfield  {title}
  {\enquote {\bibinfo {title} {{Quantum Teleportation Between Distant Matter
  Qubits}},}\ }\href {\doibase 10.1126/science.1167209} {\bibfield  {journal}
  {\bibinfo  {journal} {Science}\ }\textbf {\bibinfo {volume} {323}},\ \bibinfo
  {pages} {486--489} (\bibinfo {year} {2009})}\BibitemShut {NoStop}%
\bibitem [{\citenamefont {Lim}\ \emph {et~al.}(2018)\citenamefont {Lim},
  \citenamefont {Welinski}, \citenamefont {Ferrier}, \citenamefont {Goldner},\
  and\ \citenamefont {Morton}}]{Lim_PRB2018}%
  \BibitemOpen
  \bibfield  {author} {\bibinfo {author} {\bibfnamefont {H.-J.}\ \bibnamefont
  {Lim}}, \bibinfo {author} {\bibfnamefont {S.}~\bibnamefont {Welinski}},
  \bibinfo {author} {\bibfnamefont {A.}~\bibnamefont {Ferrier}}, \bibinfo
  {author} {\bibfnamefont {P.}~\bibnamefont {Goldner}}, \ and\ \bibinfo
  {author} {\bibfnamefont {J.~J.~L.}\ \bibnamefont {Morton}},\ }\bibfield
  {title} {\enquote {\bibinfo {title} {Coherent spin dynamics of ytterbium ions
  in yttrium orthosilicate},}\ }\href {\doibase 10.1103/PhysRevB.97.064409}
  {\bibfield  {journal} {\bibinfo  {journal} {Phys. Rev. B}\ }\textbf {\bibinfo
  {volume} {97}},\ \bibinfo {pages} {064409} (\bibinfo {year}
  {2018})}\BibitemShut {NoStop}%
\bibitem [{\citenamefont {Fagundes-Peters}\ \emph {et~al.}(2007)\citenamefont
  {Fagundes-Peters}, \citenamefont {Martynyuk}, \citenamefont {Lanstedt},
  \citenamefont {Peters}, \citenamefont {Petermann}, \citenamefont {Huber},
  \citenamefont {Basun}, \citenamefont {Laguta},\ and\ \citenamefont
  {Hofstaetter}}]{Fagundes2007}%
  \BibitemOpen
  \bibfield  {author} {\bibinfo {author} {\bibfnamefont {D.}~\bibnamefont
  {Fagundes-Peters}}, \bibinfo {author} {\bibfnamefont {N.}~\bibnamefont
  {Martynyuk}}, \bibinfo {author} {\bibfnamefont {K.}~\bibnamefont {Lanstedt}},
  \bibinfo {author} {\bibfnamefont {V.}~\bibnamefont {Peters}}, \bibinfo
  {author} {\bibfnamefont {K.}~\bibnamefont {Petermann}}, \bibinfo {author}
  {\bibfnamefont {G.}~\bibnamefont {Huber}}, \bibinfo {author} {\bibfnamefont
  {S.}~\bibnamefont {Basun}}, \bibinfo {author} {\bibfnamefont
  {V.}~\bibnamefont {Laguta}}, \ and\ \bibinfo {author} {\bibfnamefont
  {A.}~\bibnamefont {Hofstaetter}},\ }\bibfield  {title} {\enquote {\bibinfo
  {title} {{{High quantum efficiency YbAG-crystals}}},}\ }\href {\doibase
  https://doi.org/10.1016/j.jlumin.2006.08.035} {\bibfield  {journal} {\bibinfo
   {journal} {J. Lumin.}\ }\textbf {\bibinfo {volume} {125}},\ \bibinfo {pages}
  {238 -- 247} (\bibinfo {year} {2007})}\BibitemShut {NoStop}%
\bibitem [{\citenamefont {Boulon}\ \emph {et~al.}(2008)\citenamefont {Boulon},
  \citenamefont {Guyot}, \citenamefont {Canibano}, \citenamefont {Hraiech},\
  and\ \citenamefont {Yoshikawa}}]{Boulon2008}%
  \BibitemOpen
  \bibfield  {author} {\bibinfo {author} {\bibfnamefont {G.}~\bibnamefont
  {Boulon}}, \bibinfo {author} {\bibfnamefont {Y.}~\bibnamefont {Guyot}},
  \bibinfo {author} {\bibfnamefont {H.}~\bibnamefont {Canibano}}, \bibinfo
  {author} {\bibfnamefont {S.}~\bibnamefont {Hraiech}}, \ and\ \bibinfo
  {author} {\bibfnamefont {A.}~\bibnamefont {Yoshikawa}},\ }\bibfield  {title}
  {\enquote {\bibinfo {title} {{Characterization and comparison of
  Yb$^{3+}$-doped YAlO$_3$ perovskite crystals (Yb:YAP) with Yb$^{3+}$-doped
  Y$_3$Al$_5$O$_{12}$ garnet crystals (Yb:YAG) for laser application}},}\
  }\href {\doibase 10.1364/JOSAB.25.000884} {\bibfield  {journal} {\bibinfo
  {journal} {J. Opt. Soc. Am. B}\ }\textbf {\bibinfo {volume} {25}},\ \bibinfo
  {pages} {884--896} (\bibinfo {year} {2008})}\BibitemShut {NoStop}%
\bibitem [{\citenamefont {Xia}\ \emph {et~al.}(2015)\citenamefont {Xia},
  \citenamefont {Kolesov}, \citenamefont {Wang}, \citenamefont {Siyushev},
  \citenamefont {Reuter}, \citenamefont {Kornher}, \citenamefont {Kukharchyk},
  \citenamefont {Wieck}, \citenamefont {Villa}, \citenamefont {Yang},\ and\
  \citenamefont {Wrachtrup}}]{XiaPRL2015}%
  \BibitemOpen
  \bibfield  {author} {\bibinfo {author} {\bibfnamefont {K.}~\bibnamefont
  {Xia}}, \bibinfo {author} {\bibfnamefont {R.}~\bibnamefont {Kolesov}},
  \bibinfo {author} {\bibfnamefont {Ya}~\bibnamefont {Wang}}, \bibinfo {author}
  {\bibfnamefont {P.}~\bibnamefont {Siyushev}}, \bibinfo {author}
  {\bibfnamefont {R.}~\bibnamefont {Reuter}}, \bibinfo {author} {\bibfnamefont
  {T.}~\bibnamefont {Kornher}}, \bibinfo {author} {\bibfnamefont
  {N.}~\bibnamefont {Kukharchyk}}, \bibinfo {author} {\bibfnamefont {A.~D.}\
  \bibnamefont {Wieck}}, \bibinfo {author} {\bibfnamefont {B.}~\bibnamefont
  {Villa}}, \bibinfo {author} {\bibfnamefont {S.}~\bibnamefont {Yang}}, \ and\
  \bibinfo {author} {\bibfnamefont {J.}~\bibnamefont {Wrachtrup}},\ }\bibfield
  {title} {\enquote {\bibinfo {title} {{All-Optical Preparation of Coherent
  Dark States of a Single Rare Earth Ion Spin in a Crystal}},}\ }\href
  {\doibase 10.1103/PhysRevLett.115.093602} {\bibfield  {journal} {\bibinfo
  {journal} {Phys. Rev. Lett.}\ }\textbf {\bibinfo {volume} {115}},\ \bibinfo
  {pages} {093602} (\bibinfo {year} {2015})}\BibitemShut {NoStop}%
\bibitem [{\citenamefont {Bussi\`{e}res}\ \emph {et~al.}(2014)\citenamefont
  {Bussi\`{e}res}, \citenamefont {Clausen}, \citenamefont {Tiranov},
  \citenamefont {Korzh}, \citenamefont {Verma}, \citenamefont {Nam},
  \citenamefont {Marsili}, \citenamefont {Ferrier}, \citenamefont {Goldner},
  \citenamefont {Herrmann}, \citenamefont {Silberhorn}, \citenamefont {Sohler},
  \citenamefont {Afzelius},\ and\ \citenamefont {Gisin}}]{Bussieres2014}%
  \BibitemOpen
  \bibfield  {author} {\bibinfo {author} {\bibfnamefont {F.}~\bibnamefont
  {Bussi\`{e}res}}, \bibinfo {author} {\bibfnamefont {C.}~\bibnamefont
  {Clausen}}, \bibinfo {author} {\bibfnamefont {A.}~\bibnamefont {Tiranov}},
  \bibinfo {author} {\bibfnamefont {B.}~\bibnamefont {Korzh}}, \bibinfo
  {author} {\bibfnamefont {V.~B.}\ \bibnamefont {Verma}}, \bibinfo {author}
  {\bibfnamefont {S.~W.}\ \bibnamefont {Nam}}, \bibinfo {author} {\bibfnamefont
  {F.}~\bibnamefont {Marsili}}, \bibinfo {author} {\bibfnamefont
  {A.}~\bibnamefont {Ferrier}}, \bibinfo {author} {\bibfnamefont
  {P.}~\bibnamefont {Goldner}}, \bibinfo {author} {\bibfnamefont
  {H.}~\bibnamefont {Herrmann}}, \bibinfo {author} {\bibfnamefont
  {C.}~\bibnamefont {Silberhorn}}, \bibinfo {author} {\bibfnamefont
  {W.}~\bibnamefont {Sohler}}, \bibinfo {author} {\bibfnamefont
  {M.}~\bibnamefont {Afzelius}}, \ and\ \bibinfo {author} {\bibfnamefont
  {N.}~\bibnamefont {Gisin}},\ }\bibfield  {title} {\enquote {\bibinfo {title}
  {Quantum teleportation from a telecom-wavelength photon to a solid-state
  quantum memory},}\ }\href {\doibase 10.1038/nphoton.2014.215} {\bibfield
  {journal} {\bibinfo  {journal} {Nat. Photonics}\ }\textbf {\bibinfo {volume}
  {8}},\ \bibinfo {pages} {775--778} (\bibinfo {year} {2014})}\BibitemShut
  {NoStop}%
\bibitem [{\citenamefont {Saglamyurek}\ \emph {et~al.}(2015)\citenamefont
  {Saglamyurek}, \citenamefont {Jin}, \citenamefont {Verma}, \citenamefont
  {Shaw}, \citenamefont {Marsili}, \citenamefont {Nam}, \citenamefont {Oblak},\
  and\ \citenamefont {Tittel}}]{Saglamyurek2015}%
  \BibitemOpen
  \bibfield  {author} {\bibinfo {author} {\bibfnamefont {E.}~\bibnamefont
  {Saglamyurek}}, \bibinfo {author} {\bibfnamefont {J.}~\bibnamefont {Jin}},
  \bibinfo {author} {\bibfnamefont {V.~B.}\ \bibnamefont {Verma}}, \bibinfo
  {author} {\bibfnamefont {M.~D.}\ \bibnamefont {Shaw}}, \bibinfo {author}
  {\bibfnamefont {F.}~\bibnamefont {Marsili}}, \bibinfo {author} {\bibfnamefont
  {S.~W.}\ \bibnamefont {Nam}}, \bibinfo {author} {\bibfnamefont
  {D.}~\bibnamefont {Oblak}}, \ and\ \bibinfo {author} {\bibfnamefont
  {W.}~\bibnamefont {Tittel}},\ }\bibfield  {title} {\enquote {\bibinfo {title}
  {{Quantum storage of entangled telecom-wavelength photons in an erbium-doped
  optical fibre}},}\ }\href {\doibase 10.1038/nphoton.2014.311} {\bibfield
  {journal} {\bibinfo  {journal} {Nat. Photonics}\ }\textbf {\bibinfo {volume}
  {9}},\ \bibinfo {pages} {83--87} (\bibinfo {year} {2015})}\BibitemShut
  {NoStop}%
\bibitem [{\citenamefont {Zhong}\ \emph {et~al.}(2017)\citenamefont {Zhong},
  \citenamefont {Kindem}, \citenamefont {Bartholomew}, \citenamefont {Rochman},
  \citenamefont {Craiciu}, \citenamefont {Miyazono}, \citenamefont
  {Bettinelli}, \citenamefont {Cavalli}, \citenamefont {Verma}, \citenamefont
  {Nam}, \citenamefont {Marsili}, \citenamefont {Shaw}, \citenamefont {Beyer},\
  and\ \citenamefont {Faraon}}]{Zhong2017}%
  \BibitemOpen
  \bibfield  {author} {\bibinfo {author} {\bibfnamefont {T.}~\bibnamefont
  {Zhong}}, \bibinfo {author} {\bibfnamefont {J.~M.}\ \bibnamefont {Kindem}},
  \bibinfo {author} {\bibfnamefont {J.~G.}\ \bibnamefont {Bartholomew}},
  \bibinfo {author} {\bibfnamefont {J.}~\bibnamefont {Rochman}}, \bibinfo
  {author} {\bibfnamefont {I.}~\bibnamefont {Craiciu}}, \bibinfo {author}
  {\bibfnamefont {E.}~\bibnamefont {Miyazono}}, \bibinfo {author}
  {\bibfnamefont {M.}~\bibnamefont {Bettinelli}}, \bibinfo {author}
  {\bibfnamefont {E.}~\bibnamefont {Cavalli}}, \bibinfo {author} {\bibfnamefont
  {V.}~\bibnamefont {Verma}}, \bibinfo {author} {\bibfnamefont {S.~W.}\
  \bibnamefont {Nam}}, \bibinfo {author} {\bibfnamefont {F.}~\bibnamefont
  {Marsili}}, \bibinfo {author} {\bibfnamefont {M.~D.}\ \bibnamefont {Shaw}},
  \bibinfo {author} {\bibfnamefont {A.~D.}\ \bibnamefont {Beyer}}, \ and\
  \bibinfo {author} {\bibfnamefont {A.}~\bibnamefont {Faraon}},\ }\bibfield
  {title} {\enquote {\bibinfo {title} {Nanophotonic rare-earth quantum memory
  with optically controlled retrieval},}\ }\href {\doibase
  10.1126/science.aan5959} {\bibfield  {journal} {\bibinfo  {journal}
  {Science}\ }\textbf {\bibinfo {volume} {357}},\ \bibinfo {pages} {1392--1395}
  (\bibinfo {year} {2017})}\BibitemShut {NoStop}%
\bibitem [{\citenamefont {Paddison}\ \emph {et~al.}(2017)\citenamefont
  {Paddison}, \citenamefont {Daum}, \citenamefont {Dun}, \citenamefont
  {Ehlers}, \citenamefont {Liu}, \citenamefont {Stone}, \citenamefont {Zhou},\
  and\ \citenamefont {Mourigal}}]{paddison2017continuous}%
  \BibitemOpen
  \bibfield  {author} {\bibinfo {author} {\bibfnamefont {J.~A.~M.}\
  \bibnamefont {Paddison}}, \bibinfo {author} {\bibfnamefont {M.}~\bibnamefont
  {Daum}}, \bibinfo {author} {\bibfnamefont {Z.}~\bibnamefont {Dun}}, \bibinfo
  {author} {\bibfnamefont {G.}~\bibnamefont {Ehlers}}, \bibinfo {author}
  {\bibfnamefont {Y.}~\bibnamefont {Liu}}, \bibinfo {author} {\bibfnamefont
  {M.~B.}\ \bibnamefont {Stone}}, \bibinfo {author} {\bibfnamefont
  {H.}~\bibnamefont {Zhou}}, \ and\ \bibinfo {author} {\bibfnamefont
  {M.}~\bibnamefont {Mourigal}},\ }\bibfield  {title} {\enquote {\bibinfo
  {title} {{Continuous excitations of the triangular-lattice quantum spin
  liquid YbMgGaO$_4$}},}\ }\href {\doibase 10.1038/nphys3971} {\bibfield
  {journal} {\bibinfo  {journal} {Nat. Phys.}\ }\textbf {\bibinfo {volume}
  {13}},\ \bibinfo {pages} {117--122} (\bibinfo {year} {2017})}\BibitemShut
  {NoStop}%
\bibitem [{\citenamefont {Baenitz}\ \emph {et~al.}(2018)\citenamefont
  {Baenitz}, \citenamefont {Schlender}, \citenamefont {Sichelschmidt},
  \citenamefont {Onykiienko}, \citenamefont {Zangeneh}, \citenamefont
  {Ranjith}, \citenamefont {Sarkar}, \citenamefont {Hozoi}, \citenamefont
  {Walker}, \citenamefont {Orain}, \citenamefont {Yasuoka}, \citenamefont
  {van~den Brink}, \citenamefont {Klauss}, \citenamefont {Inosov},\ and\
  \citenamefont {Doert}}]{baenitz2018naybs}%
  \BibitemOpen
  \bibfield  {author} {\bibinfo {author} {\bibfnamefont {M.}~\bibnamefont
  {Baenitz}}, \bibinfo {author} {\bibfnamefont {P.}\ \bibnamefont
  {Schlender}}, \bibinfo {author} {\bibfnamefont {J.}~\bibnamefont
  {Sichelschmidt}}, \bibinfo {author} {\bibfnamefont {Y.~A.}\ \bibnamefont
  {Onykiienko}}, \bibinfo {author} {\bibfnamefont {Z.}~\bibnamefont
  {Zangeneh}}, \bibinfo {author} {\bibfnamefont {K.~M.}\ \bibnamefont
  {Ranjith}}, \bibinfo {author} {\bibfnamefont {R.}~\bibnamefont {Sarkar}},
  \bibinfo {author} {\bibfnamefont {L.}~\bibnamefont {Hozoi}}, \bibinfo
  {author} {\bibfnamefont {H.~C.}\ \bibnamefont {Walker}}, \bibinfo {author}
  {\bibfnamefont {J.-C.}\ \bibnamefont {Orain}}, \bibinfo {author}
  {\bibfnamefont {H.}~\bibnamefont {Yasuoka}}, \bibinfo {author} {\bibfnamefont
  {J.}~\bibnamefont {van~den Brink}}, \bibinfo {author} {\bibfnamefont {H.~H.}\
  \bibnamefont {Klauss}}, \bibinfo {author} {\bibfnamefont {D.~S.}\
  \bibnamefont {Inosov}}, \ and\ \bibinfo {author} {\bibfnamefont {Th.}\
  \bibnamefont {Doert}},\ }\bibfield  {title} {\enquote {\bibinfo {title}
  {{NaYbS$_2$: A planar spin-1/2 triangular-lattice magnet and putative spin
  liquid}},}\ }\href {\doibase 10.1103/PhysRevB.98.220409} {\bibfield
  {journal} {\bibinfo  {journal} {Phys. Rev. B}\ }\textbf {\bibinfo {volume}
  {98}},\ \bibinfo {pages} {220409(R)} (\bibinfo {year} {2018})}\BibitemShut
  {NoStop}%
\bibitem [{\citenamefont {Gao}\ \emph {et~al.}(2019)\citenamefont {Gao},
  \citenamefont {Chen}, \citenamefont {Tam}, \citenamefont {Huang},
  \citenamefont {Sasmal}, \citenamefont {Adroja}, \citenamefont {Ye},
  \citenamefont {Cao}, \citenamefont {Sala}, \citenamefont {Stone},
  \citenamefont {Baines}, \citenamefont {Verezhak}, \citenamefont {Hu},
  \citenamefont {Chung}, \citenamefont {Xu}, \citenamefont {Cheong},
  \citenamefont {Nallaiyan}, \citenamefont {Spagna}, \citenamefont {Maple},
  \citenamefont {Nevidomskyy}, \citenamefont {Morosan}, \citenamefont {Chen},\
  and\ \citenamefont {Dai}}]{gao2019experimental}%
  \BibitemOpen
  \bibfield  {author} {\bibinfo {author} {\bibfnamefont {B.}~\bibnamefont
  {Gao}}, \bibinfo {author} {\bibfnamefont {T.}~\bibnamefont {Chen}}, \bibinfo
  {author} {\bibfnamefont {D.~W.}\ \bibnamefont {Tam}}, \bibinfo {author}
  {\bibfnamefont {C.-L.}\ \bibnamefont {Huang}}, \bibinfo {author}
  {\bibfnamefont {K.}~\bibnamefont {Sasmal}}, \bibinfo {author} {\bibfnamefont
  {D.~T.}\ \bibnamefont {Adroja}}, \bibinfo {author} {\bibfnamefont
  {F.}~\bibnamefont {Ye}}, \bibinfo {author} {\bibfnamefont {H.}~\bibnamefont
  {Cao}}, \bibinfo {author} {\bibfnamefont {G.}~\bibnamefont {Sala}}, \bibinfo
  {author} {\bibfnamefont {M.~B.}\ \bibnamefont {Stone}}, \bibinfo {author}
  {\bibfnamefont {C.}~\bibnamefont {Baines}}, \bibinfo {author} {\bibfnamefont
  {J.~A.~T.}\ \bibnamefont {Verezhak}}, \bibinfo {author} {\bibfnamefont
  {H.}~\bibnamefont {Hu}}, \bibinfo {author} {\bibfnamefont {J.-H.}\
  \bibnamefont {Chung}}, \bibinfo {author} {\bibfnamefont {X.}~\bibnamefont
  {Xu}}, \bibinfo {author} {\bibfnamefont {S.-W.}\ \bibnamefont {Cheong}},
  \bibinfo {author} {\bibfnamefont {M.}~\bibnamefont {Nallaiyan}}, \bibinfo
  {author} {\bibfnamefont {S.}~\bibnamefont {Spagna}}, \bibinfo {author}
  {\bibfnamefont {M.~B.}\ \bibnamefont {Maple}}, \bibinfo {author}
  {\bibfnamefont {A.~H.}\ \bibnamefont {Nevidomskyy}}, \bibinfo {author}
  {\bibfnamefont {E.}~\bibnamefont {Morosan}}, \bibinfo {author} {\bibfnamefont
  {G.}~\bibnamefont {Chen}}, \ and\ \bibinfo {author} {\bibfnamefont
  {P.}~\bibnamefont {Dai}},\ }\bibfield  {title} {\enquote {\bibinfo {title}
  {{Experimental signatures of a three-dimensional quantum spin liquid in
  effective spin-1/2 Ce$_2$Zr$_2$O$_7$ pyrochlore}},}\ }\href {\doibase
  10.1038/s41567-019-0577-6} {\bibfield  {journal} {\bibinfo  {journal} {Nature
  Phys.}\ }\textbf {\bibinfo {volume} {15}},\ \bibinfo {pages} {1052--1057}
  (\bibinfo {year} {2019})}\BibitemShut {NoStop}%
\bibitem [{\citenamefont {Hester}\ \emph {et~al.}(2019)\citenamefont {Hester},
  \citenamefont {Nair}, \citenamefont {Reeder}, \citenamefont {Yahne},
  \citenamefont {DeLazzer}, \citenamefont {Berges}, \citenamefont {Ziat},
  \citenamefont {Neilson}, \citenamefont {Aczel}, \citenamefont {Sala},
  \citenamefont {Quilliam},\ and\ \citenamefont {Ross}}]{hester2019novel}%
  \BibitemOpen
  \bibfield  {author} {\bibinfo {author} {\bibfnamefont {G.}~\bibnamefont
  {Hester}}, \bibinfo {author} {\bibfnamefont {H.S.}\ \bibnamefont {Nair}},
  \bibinfo {author} {\bibfnamefont {T.}~\bibnamefont {Reeder}}, \bibinfo
  {author} {\bibfnamefont {D.~R.}\ \bibnamefont {Yahne}}, \bibinfo {author}
  {\bibfnamefont {T.~N.}\ \bibnamefont {DeLazzer}}, \bibinfo {author}
  {\bibfnamefont {L.}~\bibnamefont {Berges}}, \bibinfo {author} {\bibfnamefont
  {D.}~\bibnamefont {Ziat}}, \bibinfo {author} {\bibfnamefont {J.~R.}\
  \bibnamefont {Neilson}}, \bibinfo {author} {\bibfnamefont {A.~A.}\
  \bibnamefont {Aczel}}, \bibinfo {author} {\bibfnamefont {G.}~\bibnamefont
  {Sala}}, \bibinfo {author} {\bibfnamefont {J.~A.}\ \bibnamefont {Quilliam}},
  \ and\ \bibinfo {author} {\bibfnamefont {K.~A.}\ \bibnamefont {Ross}},\
  }\bibfield  {title} {\enquote {\bibinfo {title} {{Novel Strongly Spin-Orbit
  Coupled Quantum Dimer Magnet: Yb$_2$Si$_2$O$_7$}},}\ }\href {\doibase
  10.1103/PhysRevLett.123.027201} {\bibfield  {journal} {\bibinfo  {journal}
  {Phys. Rev. Lett.}\ }\textbf {\bibinfo {volume} {123}},\ \bibinfo {pages}
  {027201} (\bibinfo {year} {2019})}\BibitemShut {NoStop}%
\bibitem [{\citenamefont {Nikitin}\ \emph {et~al.}(2018)\citenamefont
  {Nikitin}, \citenamefont {Wu}, \citenamefont {Sefat}, \citenamefont
  {Shaykhutdinov}, \citenamefont {Lu}, \citenamefont {Meng}, \citenamefont
  {Pomjakushina}, \citenamefont {Conder}, \citenamefont {Ehlers}, \citenamefont
  {Lumsden}, \citenamefont {Kolesnikov}, \citenamefont {Barilo}, \citenamefont
  {Guretskii}, \citenamefont {Inosov},\ and\ \citenamefont
  {Podlesnyak}}]{Nikitin2018}%
  \BibitemOpen
  \bibfield  {author} {\bibinfo {author} {\bibfnamefont {S.~E.}\ \bibnamefont
  {Nikitin}}, \bibinfo {author} {\bibfnamefont {L.~S.}\ \bibnamefont {Wu}},
  \bibinfo {author} {\bibfnamefont {A.~S.}\ \bibnamefont {Sefat}}, \bibinfo
  {author} {\bibfnamefont {K.~A.}\ \bibnamefont {Shaykhutdinov}}, \bibinfo
  {author} {\bibfnamefont {Z.}~\bibnamefont {Lu}}, \bibinfo {author}
  {\bibfnamefont {S.}~\bibnamefont {Meng}}, \bibinfo {author} {\bibfnamefont
  {E.~V.}\ \bibnamefont {Pomjakushina}}, \bibinfo {author} {\bibfnamefont
  {K.}~\bibnamefont {Conder}}, \bibinfo {author} {\bibfnamefont
  {G.}~\bibnamefont {Ehlers}}, \bibinfo {author} {\bibfnamefont {M.~D.}\
  \bibnamefont {Lumsden}}, \bibinfo {author} {\bibfnamefont {A.~I.}\
  \bibnamefont {Kolesnikov}}, \bibinfo {author} {\bibfnamefont
  {S.}~\bibnamefont {Barilo}}, \bibinfo {author} {\bibfnamefont {S.~A.}\
  \bibnamefont {Guretskii}}, \bibinfo {author} {\bibfnamefont {D.~S.}\
  \bibnamefont {Inosov}}, \ and\ \bibinfo {author} {\bibfnamefont
  {A.}~\bibnamefont {Podlesnyak}},\ }\bibfield  {title} {\enquote {\bibinfo
  {title} {{Decoupled spin dynamics in the rare-earth orthoferrite
  ${\mathrm{YbFeO}}_{3}$: Evolution of magnetic excitations through the
  spin-reorientation transition}},}\ }\href {\doibase
  10.1103/PhysRevB.98.064424} {\bibfield  {journal} {\bibinfo  {journal} {Phys.
  Rev. B}\ }\textbf {\bibinfo {volume} {98}},\ \bibinfo {pages} {064424}
  (\bibinfo {year} {2018})}\BibitemShut {NoStop}%
\bibitem [{\citenamefont {Norman}(2016)}]{norman2016colloquium}%
  \BibitemOpen
  \bibfield  {author} {\bibinfo {author} {\bibfnamefont {M.~R.}\ \bibnamefont
  {Norman}},\ }\bibfield  {title} {\enquote {\bibinfo {title} {Colloquium:
  Herbertsmithite and the search for the quantum spin liquid},}\ }\href
  {\doibase 10.1103/RevModPhys.88.041002} {\bibfield  {journal} {\bibinfo
  {journal} {Rev. Mod. Phys.}\ }\textbf {\bibinfo {volume} {88}},\ \bibinfo
  {pages} {041002} (\bibinfo {year} {2016})}\BibitemShut {NoStop}%
\bibitem [{\citenamefont {Stone}\ \emph {et~al.}(2006)\citenamefont {Stone},
  \citenamefont {Zaliznyak}, \citenamefont {Hong}, \citenamefont {Broholm},\
  and\ \citenamefont {Reich}}]{Stone2006}%
  \BibitemOpen
  \bibfield  {author} {\bibinfo {author} {\bibfnamefont {M.~B.}\ \bibnamefont
  {Stone}}, \bibinfo {author} {\bibfnamefont {I.~A.}\ \bibnamefont
  {Zaliznyak}}, \bibinfo {author} {\bibfnamefont {T.}~\bibnamefont {Hong}},
  \bibinfo {author} {\bibfnamefont {C.~L.}\ \bibnamefont {Broholm}}, \ and\
  \bibinfo {author} {\bibfnamefont {D.~H.}\ \bibnamefont {Reich}},\ }\bibfield
  {title} {\enquote {\bibinfo {title} {Quasiparticle breakdown in a quantum
  spin liquid},}\ }\href {\doibase 10.1038/nature04593} {\bibfield  {journal}
  {\bibinfo  {journal} {Nature}\ }\textbf {\bibinfo {volume} {440}},\ \bibinfo
  {pages} {187--190} (\bibinfo {year} {2006})}\BibitemShut {NoStop}%
\bibitem [{\citenamefont {Bastien}\ \emph {et~al.}(2015)\citenamefont
  {Bastien}, \citenamefont {Mourigal}, \citenamefont {Christensen},
  \citenamefont {Nilsen}, \citenamefont {Tregenna-Piggott}, \citenamefont
  {Perring}, \citenamefont {Enderle}, \citenamefont {McMorrow}, \citenamefont
  {Ivanov},\ and\ \citenamefont {R{\o}nnow}}]{dalla2015fractional}%
  \BibitemOpen
  \bibfield  {author} {\bibinfo {author} {\bibfnamefont {D.~P.}\ \bibnamefont
  {Bastien}}, \bibinfo {author} {\bibfnamefont {M.}~\bibnamefont {Mourigal}},
  \bibinfo {author} {\bibfnamefont {N.~B.}\ \bibnamefont {Christensen}},
  \bibinfo {author} {\bibfnamefont {G.~J.}\ \bibnamefont {Nilsen}}, \bibinfo
  {author} {\bibfnamefont {P.}~\bibnamefont {Tregenna-Piggott}}, \bibinfo
  {author} {\bibfnamefont {T.~G.}\ \bibnamefont {Perring}}, \bibinfo {author}
  {\bibfnamefont {M.}~\bibnamefont {Enderle}}, \bibinfo {author} {\bibfnamefont
  {D.~F.}\ \bibnamefont {McMorrow}}, \bibinfo {author} {\bibfnamefont {D.~A.}\
  \bibnamefont {Ivanov}}, \ and\ \bibinfo {author} {\bibfnamefont {H.~M.}\
  \bibnamefont {R{\o}nnow}},\ }\bibfield  {title} {\enquote {\bibinfo {title}
  {Fractional excitations in the square-lattice quantum antiferromagnet},}\
  }\href {\doibase 10.1038/NPHYS3172} {\bibfield  {journal} {\bibinfo
  {journal} {Nat. Phys.}\ }\textbf {\bibinfo {volume} {11}},\ \bibinfo {pages}
  {62--68} (\bibinfo {year} {2015})}\BibitemShut {NoStop}%
\bibitem [{\citenamefont {Kim}\ \emph {et~al.}(1996)\citenamefont {Kim},
  \citenamefont {Matsuura}, \citenamefont {Shen}, \citenamefont {Motoyama},
  \citenamefont {Eisaki}, \citenamefont {Uchida}, \citenamefont {Tohyama},\
  and\ \citenamefont {Maekawa}}]{kim1996observation}%
  \BibitemOpen
  \bibfield  {author} {\bibinfo {author} {\bibfnamefont {C.}~\bibnamefont
  {Kim}}, \bibinfo {author} {\bibfnamefont {A.~Y.}\ \bibnamefont {Matsuura}},
  \bibinfo {author} {\bibfnamefont {Z.-X.}\ \bibnamefont {Shen}}, \bibinfo
  {author} {\bibfnamefont {N.}~\bibnamefont {Motoyama}}, \bibinfo {author}
  {\bibfnamefont {H.}~\bibnamefont {Eisaki}}, \bibinfo {author} {\bibfnamefont
  {S.}~\bibnamefont {Uchida}}, \bibinfo {author} {\bibfnamefont
  {T.}~\bibnamefont {Tohyama}}, \ and\ \bibinfo {author} {\bibfnamefont
  {S.}~\bibnamefont {Maekawa}},\ }\bibfield  {title} {\enquote {\bibinfo
  {title} {{Observation of spin-charge separation in one-dimensional
  SrCuO$_2$}},}\ }\href {\doibase 10.1103/PhysRevLett.77.4054} {\bibfield
  {journal} {\bibinfo  {journal} {Phys. Rev. Lett.}\ }\textbf {\bibinfo
  {volume} {77}},\ \bibinfo {pages} {4054} (\bibinfo {year}
  {1996})}\BibitemShut {NoStop}%
\bibitem [{\citenamefont {Motoyama}\ \emph {et~al.}(1996)\citenamefont
  {Motoyama}, \citenamefont {Eisaki},\ and\ \citenamefont
  {Uchida}}]{motoyama1996magnetic}%
  \BibitemOpen
  \bibfield  {author} {\bibinfo {author} {\bibfnamefont {N.}~\bibnamefont
  {Motoyama}}, \bibinfo {author} {\bibfnamefont {H.}~\bibnamefont {Eisaki}}, \
  and\ \bibinfo {author} {\bibfnamefont {S.}~\bibnamefont {Uchida}},\
  }\bibfield  {title} {\enquote {\bibinfo {title} {{Magnetic susceptibility of
  ideal spin-1/2 Heisenberg antiferromagnetic chain systems, Sr$_2$CuO$_3$ and
  SrCuO$_2$}},}\ }\href {\doibase 10.1103/PhysRevLett.76.3212} {\bibfield
  {journal} {\bibinfo  {journal} {Phys. Rev. Lett.}\ }\textbf {\bibinfo
  {volume} {76}},\ \bibinfo {pages} {3212} (\bibinfo {year}
  {1996})}\BibitemShut {NoStop}%
\bibitem [{\citenamefont {Zaliznyak}\ \emph {et~al.}(2004)\citenamefont
  {Zaliznyak}, \citenamefont {Woo}, \citenamefont {Perring}, \citenamefont
  {Broholm}, \citenamefont {Frost},\ and\ \citenamefont
  {Takagi}}]{Zaliznyak_PRL2004}%
  \BibitemOpen
  \bibfield  {author} {\bibinfo {author} {\bibfnamefont {I.~A.}\ \bibnamefont
  {Zaliznyak}}, \bibinfo {author} {\bibfnamefont {H.}~\bibnamefont {Woo}},
  \bibinfo {author} {\bibfnamefont {T.~G.}\ \bibnamefont {Perring}}, \bibinfo
  {author} {\bibfnamefont {C.~L.}\ \bibnamefont {Broholm}}, \bibinfo {author}
  {\bibfnamefont {C.~D.}\ \bibnamefont {Frost}}, \ and\ \bibinfo {author}
  {\bibfnamefont {H.}~\bibnamefont {Takagi}},\ }\bibfield  {title} {\enquote
  {\bibinfo {title} {{Spinons in the Strongly Correlated Copper Oxide Chains in
  ${\mathrm{S}\mathrm{r}\mathrm{C}\mathrm{u}\mathrm{O}}_{2}$}},}\ }\href
  {\doibase 10.1103/PhysRevLett.93.087202} {\bibfield  {journal} {\bibinfo
  {journal} {Phys. Rev. Lett.}\ }\textbf {\bibinfo {volume} {93}},\ \bibinfo
  {pages} {087202} (\bibinfo {year} {2004})}\BibitemShut {NoStop}%
\bibitem [{\citenamefont {Walters}\ \emph {et~al.}(2009)\citenamefont
  {Walters}, \citenamefont {Perring}, \citenamefont {Caux}, \citenamefont
  {Savici}, \citenamefont {Gu}, \citenamefont {Lee}, \citenamefont {Ku},\ and\
  \citenamefont {Zaliznyak}}]{Walters_NaturePhysics2009}%
  \BibitemOpen
  \bibfield  {author} {\bibinfo {author} {\bibfnamefont {A.~C.}\ \bibnamefont
  {Walters}}, \bibinfo {author} {\bibfnamefont {T.~G.}\ \bibnamefont
  {Perring}}, \bibinfo {author} {\bibfnamefont {J.-S.}\ \bibnamefont {Caux}},
  \bibinfo {author} {\bibfnamefont {A.~T.}\ \bibnamefont {Savici}}, \bibinfo
  {author} {\bibfnamefont {G.~D.}\ \bibnamefont {Gu}}, \bibinfo {author}
  {\bibfnamefont {C.-C.}\ \bibnamefont {Lee}}, \bibinfo {author} {\bibfnamefont
  {W.}~\bibnamefont {Ku}}, \ and\ \bibinfo {author} {\bibfnamefont {I.~A.}\
  \bibnamefont {Zaliznyak}},\ }\bibfield  {title} {\enquote {\bibinfo {title}
  {Effect of covalent bonding on magnetism and the missing neutron intensity in
  copper oxide compounds},}\ }\href@noop {} {\bibfield  {journal} {\bibinfo
  {journal} {Nature Physics}\ }\textbf {\bibinfo {volume} {5}},\ \bibinfo
  {pages} {867 -- 872} (\bibinfo {year} {2009})}\BibitemShut {NoStop}%
\bibitem [{SM()}]{SM}%
  \BibitemOpen
  \href@noop {} {}\bibinfo {howpublished}
  {https://www.scientificmaterials.com}\BibitemShut {NoStop}%
\bibitem [{\citenamefont {Ehlers}\ \emph {et~al.}(2011)\citenamefont {Ehlers},
  \citenamefont {Podlesnyak}, \citenamefont {Niedziela}, \citenamefont
  {Iverson},\ and\ \citenamefont {Sokol}}]{CNCS1}%
  \BibitemOpen
  \bibfield  {author} {\bibinfo {author} {\bibfnamefont {G.}~\bibnamefont
  {Ehlers}}, \bibinfo {author} {\bibfnamefont {A.}~\bibnamefont {Podlesnyak}},
  \bibinfo {author} {\bibfnamefont {J.~L.}\ \bibnamefont {Niedziela}}, \bibinfo
  {author} {\bibfnamefont {E.~B.}\ \bibnamefont {Iverson}}, \ and\ \bibinfo
  {author} {\bibfnamefont {P.~E.}\ \bibnamefont {Sokol}},\ }\bibfield  {title}
  {\enquote {\bibinfo {title} {The new cold neutron chopper spectrometer at the
  spallation neutron source: design and performance},}\ }\href {\doibase
  10.1063/1.3626935} {\bibfield  {journal} {\bibinfo  {journal} {Rev. Sci.
  Instrum.}\ }\textbf {\bibinfo {volume} {82}},\ \bibinfo {pages} {085108}
  (\bibinfo {year} {2011})}\BibitemShut {NoStop}%
\bibitem [{\citenamefont {Ehlers}\ \emph {et~al.}(2016)\citenamefont {Ehlers},
  \citenamefont {Podlesnyak},\ and\ \citenamefont {Kolesnikov}}]{CNCS2}%
  \BibitemOpen
  \bibfield  {author} {\bibinfo {author} {\bibfnamefont {G.}~\bibnamefont
  {Ehlers}}, \bibinfo {author} {\bibfnamefont {A.}~\bibnamefont {Podlesnyak}},
  \ and\ \bibinfo {author} {\bibfnamefont {A.~I.}\ \bibnamefont {Kolesnikov}},\
  }\bibfield  {title} {\enquote {\bibinfo {title} {{The cold neutron chopper
  spectrometer at the Spallation Neutron Source - A review of the first 8 years
  of operation}},}\ }\href {\doibase 10.1063/1.4962024} {\bibfield  {journal}
  {\bibinfo  {journal} {Rev. Sci. Instrum.}\ }\textbf {\bibinfo {volume}
  {87}},\ \bibinfo {pages} {093902} (\bibinfo {year} {2016})}\BibitemShut
  {NoStop}%
\bibitem [{\citenamefont {Ewings}\ \emph {et~al.}(2016)\citenamefont {Ewings},
  \citenamefont {Buts}, \citenamefont {Le}, \citenamefont {van Duijn},
  \citenamefont {Bustinduy},\ and\ \citenamefont {Perring}}]{Horace}%
  \BibitemOpen
  \bibfield  {author} {\bibinfo {author} {\bibfnamefont {R.~A.}\ \bibnamefont
  {Ewings}}, \bibinfo {author} {\bibfnamefont {A.}~\bibnamefont {Buts}},
  \bibinfo {author} {\bibfnamefont {M.~D.}\ \bibnamefont {Le}}, \bibinfo
  {author} {\bibfnamefont {J.}~\bibnamefont {van Duijn}}, \bibinfo {author}
  {\bibfnamefont {I.}~\bibnamefont {Bustinduy}}, \ and\ \bibinfo {author}
  {\bibfnamefont {T.~G.}\ \bibnamefont {Perring}},\ }\bibfield  {title}
  {\enquote {\bibinfo {title} {{HORACE: software for the analysis of data from
  single crystal spectroscopy experiments at time-of-flight neutron
  instruments}},}\ }\href {\doibase 10.1016/j.nima.2016.07.036} {\bibfield
  {journal} {\bibinfo  {journal} {Nucl. Instrum. Methods Phys. Res. Sect. A}\
  }\textbf {\bibinfo {volume} {834}},\ \bibinfo {pages} {3132--142} (\bibinfo
  {year} {2016})}\BibitemShut {NoStop}%
\bibitem [{\citenamefont {Arnold}\ \emph {et~al.}(2014)\citenamefont {Arnold},
  \citenamefont {Bilheux}, \citenamefont {Borreguero}, \citenamefont {Buts},
  \citenamefont {Campbell}, \citenamefont {Chapon}, \citenamefont {Doucet},
  \citenamefont {Draper}, \citenamefont {Leal}, \citenamefont {Gigg},
  \citenamefont {Lynch}, \citenamefont {Markvardsen}, \citenamefont
  {Mikkelson}, \citenamefont {Mikkelson}, \citenamefont {Miller}, \citenamefont
  {Palmen}, \citenamefont {Parker}, \citenamefont {Passos}, \citenamefont
  {Perring}, \citenamefont {Peterson}, \citenamefont {Ren}, \citenamefont
  {Reuter}, \citenamefont {Savici}, \citenamefont {Taylor}, \citenamefont
  {Taylor}, \citenamefont {Tolchenov}, \citenamefont {Zhou},\ and\
  \citenamefont {Zikovsky}}]{Mantid}%
  \BibitemOpen
  \bibfield  {author} {\bibinfo {author} {\bibfnamefont {O.}~\bibnamefont
  {Arnold}}, \bibinfo {author} {\bibfnamefont {J.~C.}\ \bibnamefont {Bilheux}},
  \bibinfo {author} {\bibfnamefont {J.~M.}\ \bibnamefont {Borreguero}},
  \bibinfo {author} {\bibfnamefont {A.}~\bibnamefont {Buts}}, \bibinfo {author}
  {\bibfnamefont {S.~I.}\ \bibnamefont {Campbell}}, \bibinfo {author}
  {\bibfnamefont {L.}~\bibnamefont {Chapon}}, \bibinfo {author} {\bibfnamefont
  {M.}~\bibnamefont {Doucet}}, \bibinfo {author} {\bibfnamefont
  {N.}~\bibnamefont {Draper}}, \bibinfo {author} {\bibfnamefont {R.~Ferraz}\
  \bibnamefont {Leal}}, \bibinfo {author} {\bibfnamefont {M.~A.}\ \bibnamefont
  {Gigg}}, \bibinfo {author} {\bibfnamefont {V.~E.}\ \bibnamefont {Lynch}},
  \bibinfo {author} {\bibfnamefont {A.}~\bibnamefont {Markvardsen}}, \bibinfo
  {author} {\bibfnamefont {D.~J.}\ \bibnamefont {Mikkelson}}, \bibinfo {author}
  {\bibfnamefont {R.~L.}\ \bibnamefont {Mikkelson}}, \bibinfo {author}
  {\bibfnamefont {R.}~\bibnamefont {Miller}}, \bibinfo {author} {\bibfnamefont
  {K.}~\bibnamefont {Palmen}}, \bibinfo {author} {\bibfnamefont
  {P.}~\bibnamefont {Parker}}, \bibinfo {author} {\bibfnamefont
  {G.}~\bibnamefont {Passos}}, \bibinfo {author} {\bibfnamefont {T.~G.}\
  \bibnamefont {Perring}}, \bibinfo {author} {\bibfnamefont {P.~F.}\
  \bibnamefont {Peterson}}, \bibinfo {author} {\bibfnamefont {S.}~\bibnamefont
  {Ren}}, \bibinfo {author} {\bibfnamefont {M.~A.}\ \bibnamefont {Reuter}},
  \bibinfo {author} {\bibfnamefont {A.~T.}\ \bibnamefont {Savici}}, \bibinfo
  {author} {\bibfnamefont {J.~W.}\ \bibnamefont {Taylor}}, \bibinfo {author}
  {\bibfnamefont {R.~J.}\ \bibnamefont {Taylor}}, \bibinfo {author}
  {\bibfnamefont {R.}~\bibnamefont {Tolchenov}}, \bibinfo {author}
  {\bibfnamefont {W.}~\bibnamefont {Zhou}}, \ and\ \bibinfo {author}
  {\bibfnamefont {J.}~\bibnamefont {Zikovsky}},\ }\bibfield  {title} {\enquote
  {\bibinfo {title} {{Mantid -- Data analysis and visualization package for
  neutron scattering and $\mu$SR experiments}},}\ }\href {\doibase
  10.1016/j.nima.2014.07.029} {\bibfield  {journal} {\bibinfo  {journal} {Nucl.
  Instrum. Methods Phys. Res. Sect. A}\ }\textbf {\bibinfo {volume} {764}},\
  \bibinfo {pages} {156} (\bibinfo {year} {2014})}\BibitemShut {NoStop}%
\bibitem [{\citenamefont {Buryy}\ \emph {et~al.}(2010)\citenamefont {Buryy},
  \citenamefont {Zhydachevskii}, \citenamefont {Vasylechko}, \citenamefont
  {Sugak}, \citenamefont {Martynyuk}, \citenamefont {Ubizskii},\ and\
  \citenamefont {Becker}}]{Buryy2010}%
  \BibitemOpen
  \bibfield  {author} {\bibinfo {author} {\bibfnamefont {O.}~\bibnamefont
  {Buryy}}, \bibinfo {author} {\bibfnamefont {Ya.}\ \bibnamefont
  {Zhydachevskii}}, \bibinfo {author} {\bibfnamefont {L.}~\bibnamefont
  {Vasylechko}}, \bibinfo {author} {\bibfnamefont {D.}~\bibnamefont {Sugak}},
  \bibinfo {author} {\bibfnamefont {N.}~\bibnamefont {Martynyuk}}, \bibinfo
  {author} {\bibfnamefont {S.}~\bibnamefont {Ubizskii}}, \ and\ \bibinfo
  {author} {\bibfnamefont {K.~D.}\ \bibnamefont {Becker}},\ }\bibfield  {title}
  {\enquote {\bibinfo {title} {{Thermal changes of the crystal structure and
  the influence of thermo-chemical annealing on the optical properties of
  YbAlO$_3$ crystals}},}\ }\href {\doibase 10.1088/0953-8984/22/5/055902}
  {\bibfield  {journal} {\bibinfo  {journal} {J. Phys. Condens. Matter}\
  }\textbf {\bibinfo {volume} {22}},\ \bibinfo {pages} {055902} (\bibinfo
  {year} {2010})}\BibitemShut {NoStop}%
\bibitem [{\citenamefont {Noginov}\ \emph {et~al.}(2001)\citenamefont
  {Noginov}, \citenamefont {Loutts}, \citenamefont {Ross}, \citenamefont
  {Grandy}, \citenamefont {Noginova}, \citenamefont {Lucas},\ and\
  \citenamefont {Mapp}}]{Noginov2001}%
  \BibitemOpen
  \bibfield  {author} {\bibinfo {author} {\bibfnamefont {M.~A.}\ \bibnamefont
  {Noginov}}, \bibinfo {author} {\bibfnamefont {G.~B.}\ \bibnamefont {Loutts}},
  \bibinfo {author} {\bibfnamefont {K.}~\bibnamefont {Ross}}, \bibinfo {author}
  {\bibfnamefont {T.}~\bibnamefont {Grandy}}, \bibinfo {author} {\bibfnamefont
  {N.}~\bibnamefont {Noginova}}, \bibinfo {author} {\bibfnamefont {B.~D.}\
  \bibnamefont {Lucas}}, \ and\ \bibinfo {author} {\bibfnamefont
  {T.}~\bibnamefont {Mapp}},\ }\bibfield  {title} {\enquote {\bibinfo {title}
  {{Role of traps in photocoloration of Mn: YAlO$_3$}},}\ }\href {\doibase
  10.1364/JOSAB.18.000931} {\bibfield  {journal} {\bibinfo  {journal} {J. Opt.
  Soc. Am. B}\ }\textbf {\bibinfo {volume} {18}},\ \bibinfo {pages} {931--941}
  (\bibinfo {year} {2001})}\BibitemShut {NoStop}%
\bibitem [{\citenamefont {Furrer}\ and\ \citenamefont
  {Waldmann}(2013)}]{furrer2013magnetic}%
  \BibitemOpen
  \bibfield  {author} {\bibinfo {author} {\bibfnamefont {A.}~\bibnamefont
  {Furrer}}\ and\ \bibinfo {author} {\bibfnamefont {O.}~\bibnamefont
  {Waldmann}},\ }\bibfield  {title} {\enquote {\bibinfo {title} {Magnetic
  cluster excitations},}\ }\href {\doibase 10.1103/RevModPhys.85.367}
  {\bibfield  {journal} {\bibinfo  {journal} {Rev. Mod. Phys.}\ }\textbf
  {\bibinfo {volume} {85}},\ \bibinfo {pages} {367} (\bibinfo {year}
  {2013})}\BibitemShut {NoStop}%
\bibitem [{\citenamefont {Leuenberger}\ \emph {et~al.}(1984)\citenamefont
  {Leuenberger}, \citenamefont {Stebler}, \citenamefont {G\"udel},
  \citenamefont {Furrer}, \citenamefont {Feile},\ and\ \citenamefont
  {Kjems}}]{Leuenberger_PRB1984}%
  \BibitemOpen
  \bibfield  {author} {\bibinfo {author} {\bibfnamefont {B.}~\bibnamefont
  {Leuenberger}}, \bibinfo {author} {\bibfnamefont {A.}~\bibnamefont
  {Stebler}}, \bibinfo {author} {\bibfnamefont {H.~U.}\ \bibnamefont
  {G\"udel}}, \bibinfo {author} {\bibfnamefont {A.}~\bibnamefont {Furrer}},
  \bibinfo {author} {\bibfnamefont {R.}~\bibnamefont {Feile}}, \ and\ \bibinfo
  {author} {\bibfnamefont {J.~K.}\ \bibnamefont {Kjems}},\ }\bibfield  {title}
  {\enquote {\bibinfo {title} {Spin dynamics of an isotropic
  singlet-ground-state antiferromagnet with alternating strong and weak
  interactions: An inelastic-neutron-scattering study of the dimer compound
  ${\mathrm{cs}}_{3}$${\mathrm{cr}}_{2}$${\mathrm{br}}_{9}$},}\ }\href
  {\doibase 10.1103/PhysRevB.30.6300} {\bibfield  {journal} {\bibinfo
  {journal} {Phys. Rev. B}\ }\textbf {\bibinfo {volume} {30}},\ \bibinfo
  {pages} {6300--6307} (\bibinfo {year} {1984})}\BibitemShut {NoStop}%
\bibitem [{\citenamefont {Zheludev}\ \emph {et~al.}(1996)\citenamefont
  {Zheludev}, \citenamefont {Shirane}, \citenamefont {Sasago}, \citenamefont
  {Hase},\ and\ \citenamefont {Uchinokura}}]{Zheludev_PRB1996}%
  \BibitemOpen
  \bibfield  {author} {\bibinfo {author} {\bibfnamefont {A.}~\bibnamefont
  {Zheludev}}, \bibinfo {author} {\bibfnamefont {G.}~\bibnamefont {Shirane}},
  \bibinfo {author} {\bibfnamefont {Y.}~\bibnamefont {Sasago}}, \bibinfo
  {author} {\bibfnamefont {M.}~\bibnamefont {Hase}}, \ and\ \bibinfo {author}
  {\bibfnamefont {K.}~\bibnamefont {Uchinokura}},\ }\bibfield  {title}
  {\enquote {\bibinfo {title} {{Dimerized ground state and magnetic excitations
  in ${\mathrm{CaCuGe}}_{2}{\mathrm{O}}_{6}$}},}\ }\href {\doibase
  10.1103/PhysRevB.53.11642} {\bibfield  {journal} {\bibinfo  {journal} {Phys.
  Rev. B}\ }\textbf {\bibinfo {volume} {53}},\ \bibinfo {pages} {11642--11646}
  (\bibinfo {year} {1996})}\BibitemShut {NoStop}%
\bibitem [{Note1()}]{Note1}%
  \BibitemOpen
  \bibinfo {note} {Trimer contribution is below 1~\% of the in-filed specific
  heat, which is beyond the precision of our measurements.}\BibitemShut {Stop}%
\bibitem [{\citenamefont {Wu}\ \emph {et~al.}(2017)\citenamefont {Wu},
  \citenamefont {Nikitin}, \citenamefont {Frontzek}, \citenamefont
  {Kolesnikov}, \citenamefont {Ehlers}, \citenamefont {Lumsden}, \citenamefont
  {Shaykhutdinov}, \citenamefont {Guo}, \citenamefont {Savici}, \citenamefont
  {Gai}, \citenamefont {Sefat},\ and\ \citenamefont {Podlesnyak}}]{Wu2017}%
  \BibitemOpen
  \bibfield  {author} {\bibinfo {author} {\bibfnamefont {L.~S.}\ \bibnamefont
  {Wu}}, \bibinfo {author} {\bibfnamefont {S.~E.}\ \bibnamefont {Nikitin}},
  \bibinfo {author} {\bibfnamefont {M.}~\bibnamefont {Frontzek}}, \bibinfo
  {author} {\bibfnamefont {A.~I.}\ \bibnamefont {Kolesnikov}}, \bibinfo
  {author} {\bibfnamefont {G.}~\bibnamefont {Ehlers}}, \bibinfo {author}
  {\bibfnamefont {M.~D.}\ \bibnamefont {Lumsden}}, \bibinfo {author}
  {\bibfnamefont {K.~A.}\ \bibnamefont {Shaykhutdinov}}, \bibinfo {author}
  {\bibfnamefont {E-J.}\ \bibnamefont {Guo}}, \bibinfo {author} {\bibfnamefont
  {A.~T.}\ \bibnamefont {Savici}}, \bibinfo {author} {\bibfnamefont
  {Z.}~\bibnamefont {Gai}}, \bibinfo {author} {\bibfnamefont {A.~S.}\
  \bibnamefont {Sefat}}, \ and\ \bibinfo {author} {\bibfnamefont
  {A.}~\bibnamefont {Podlesnyak}},\ }\bibfield  {title} {\enquote {\bibinfo
  {title} {{Magnetic ground state of the Ising-like antiferromagnet
  DyScO$_3$}},}\ }\href {\doibase https://doi.org/10.1103/PhysRevB.96.144407}
  {\bibfield  {journal} {\bibinfo  {journal} {Phys. Rev. B}\ }\textbf {\bibinfo
  {volume} {96}},\ \bibinfo {pages} {144407} (\bibinfo {year}
  {2017})}\BibitemShut {NoStop}%
\bibitem [{\citenamefont {Squires}(1978)}]{Squires_book_1978}%
  \BibitemOpen
  \bibfield  {author} {\bibinfo {author} {\bibfnamefont {G.~L.}\ \bibnamefont
  {Squires}},\ }\href@noop {} {\emph {\bibinfo {title} {Introduction to the
  Theory of Thermal Neutron Scattering}}},\ \bibinfo {edition} {2012th}\ ed.\
  (\bibinfo  {publisher} {Cambridge University Press},\ \bibinfo {address}
  {England},\ \bibinfo {year} {1978})\BibitemShut {NoStop}%
\bibitem [{\citenamefont {Lovesey}(1984)}]{lovesey1984theory}%
  \BibitemOpen
  \bibfield  {author} {\bibinfo {author} {\bibfnamefont {Stephen~W}\
  \bibnamefont {Lovesey}},\ }\enquote {\bibinfo {title} {Theory of neutron
  scattering from condensed matter},}\ \ (\bibinfo  {publisher} {Oxford Science
  Publishers},\ \bibinfo {year} {1984})\BibitemShut {NoStop}%
\bibitem [{\citenamefont {Jensen}\ and\ \citenamefont
  {Mackintosh}(1991)}]{JensenMackintosh_book1991}%
  \BibitemOpen
  \bibfield  {author} {\bibinfo {author} {\bibfnamefont {J.}~\bibnamefont
  {Jensen}}\ and\ \bibinfo {author} {\bibfnamefont {A.R.}\ \bibnamefont
  {Mackintosh}},\ }\href@noop {} {\emph {\bibinfo {title} {Rare earth
  magnetism: structures and excitations}}},\ International series of monographs
  on physics\ (\bibinfo  {publisher} {Clarendon Press},\ \bibinfo {year}
  {1991})\BibitemShut {NoStop}%
\bibitem [{\citenamefont {Zaliznyak}\ and\ \citenamefont
  {Lee}(2005)}]{ZaliznyakLee_MNSChapter}%
  \BibitemOpen
  \bibfield  {author} {\bibinfo {author} {\bibfnamefont {I.~A.}\ \bibnamefont
  {Zaliznyak}}\ and\ \bibinfo {author} {\bibfnamefont {S.-H.}\ \bibnamefont
  {Lee}},\ }\bibfield  {title} {\enquote {\bibinfo {title} {Magnetic neutron
  scattering},}\ }in\ \href {\doibase 10.1007/0-387-23395-4_1} {\emph {\bibinfo
  {booktitle} {Modern Techniques for Characterizing Magnetic Materials}}},\
  \bibinfo {editor} {edited by\ \bibinfo {editor} {\bibfnamefont {Yimei}\
  \bibnamefont {Zhu}}}\ (\bibinfo  {publisher} {Springer US},\ \bibinfo {year}
  {2005})\ pp.\ \bibinfo {pages} {3--64}\BibitemShut {NoStop}%
\bibitem [{\citenamefont {Kenzelmann}\ \emph {et~al.}(2003)\citenamefont
  {Kenzelmann}, \citenamefont {Xu}, \citenamefont {Zaliznyak}, \citenamefont
  {Broholm}, \citenamefont {DiTusa}, \citenamefont {Aeppli}, \citenamefont
  {Ito}, \citenamefont {Oka},\ and\ \citenamefont
  {Takagi}}]{Kenzelmann_PRL2003}%
  \BibitemOpen
  \bibfield  {author} {\bibinfo {author} {\bibfnamefont {M.}~\bibnamefont
  {Kenzelmann}}, \bibinfo {author} {\bibfnamefont {G.}~\bibnamefont {Xu}},
  \bibinfo {author} {\bibfnamefont {I.~A.}\ \bibnamefont {Zaliznyak}}, \bibinfo
  {author} {\bibfnamefont {C.}~\bibnamefont {Broholm}}, \bibinfo {author}
  {\bibfnamefont {J.~F.}\ \bibnamefont {DiTusa}}, \bibinfo {author}
  {\bibfnamefont {G.}~\bibnamefont {Aeppli}}, \bibinfo {author} {\bibfnamefont
  {T.}~\bibnamefont {Ito}}, \bibinfo {author} {\bibfnamefont {K.}~\bibnamefont
  {Oka}}, \ and\ \bibinfo {author} {\bibfnamefont {H.}~\bibnamefont {Takagi}},\
  }\bibfield  {title} {\enquote {\bibinfo {title} {Structure of end states for
  a haldane spin chain},}\ }\href {\doibase 10.1103/PhysRevLett.90.087202}
  {\bibfield  {journal} {\bibinfo  {journal} {Phys. Rev. Lett.}\ }\textbf
  {\bibinfo {volume} {90}},\ \bibinfo {pages} {087202} (\bibinfo {year}
  {2003})}\BibitemShut {NoStop}%
\bibitem [{Note2()}]{Note2}%
  \BibitemOpen
  \bibinfo {note} {In real structure $\protect \mathbf {r}_{12} \approx (\delta
  _a, \delta _b, \protect \frac {1}{2})$, with $\delta _a \approx \pm 0.03$ and
  $\delta _b \approx \pm 0.13$. This causes an additional intensity modulation
  $I(\protect \mathbf {Q}) \propto \protect \text {cos}(2\pi \delta _a)\cdot
  \protect \text {cos}(2\pi \delta _b)$. E.g. it is responsible for the second
  ``shadow mode'' in YbFeO$_3$ spectrum~\cite {Nikitin2018}. However, it has
  only a minor influence at our range of $\protect \mathbf {Q}$.}\BibitemShut
  {Stop}%
\bibitem [{\citenamefont {Haraldsen}\ \emph {et~al.}(2005)\citenamefont
  {Haraldsen}, \citenamefont {Barnes},\ and\ \citenamefont
  {Musfeldt}}]{haraldsen2005neutron}%
  \BibitemOpen
  \bibfield  {author} {\bibinfo {author} {\bibfnamefont {J.~T.}\ \bibnamefont
  {Haraldsen}}, \bibinfo {author} {\bibfnamefont {T.}~\bibnamefont {Barnes}}, \
  and\ \bibinfo {author} {\bibfnamefont {J.~L.}\ \bibnamefont {Musfeldt}},\
  }\bibfield  {title} {\enquote {\bibinfo {title} {Neutron scattering and
  magnetic observables for $s~=~1/2$ spin clusters and molecular magnets},}\
  }\href {\doibase 10.1103/PhysRevB.71.064403} {\bibfield  {journal} {\bibinfo
  {journal} {Phys. Rev. B}\ }\textbf {\bibinfo {volume} {71}},\ \bibinfo
  {pages} {064403} (\bibinfo {year} {2005})}\BibitemShut {NoStop}%
\bibitem [{Note3()}]{Note3}%
  \BibitemOpen
  \bibinfo {note} {To increase the signal-to-noise ratio we symmetrized the
  data according to the crystal symmetry and then unfolded it back, which means
  that data at negative and positive $K$ in Fig.~\ref {Neutron_slice}~(a) are
  equivalent}\BibitemShut {NoStop}%
\end{thebibliography}

%

\end{document}